\documentclass[12pt,a4paper]{article} 

\usepackage{graphicx}

\usepackage{epsfig}
\usepackage{cite}
\usepackage{color} 
\usepackage{amssymb}
\usepackage{amsmath}
\usepackage{authblk}
\usepackage{doi}
\usepackage{dsfont}
\usepackage{setspace}

\usepackage{verbatim}

 \usepackage{slashed}
\usepackage[normalem]{ulem} 

\newcommand{\mr}{\mathrm}
\newcommand{\mc}{\mathcal}

\newcommand{\wpwpjj}{W^+W^+jj}
\newcommand{\wpzjj}{W^+Zjj}
\newcommand{\wpwmjj}{W^+W^-jj}
\newcommand{\zzjj}{ZZjj}

\newcommand{\muf}{\mu_\mr{F}}
\newcommand{\mur}{\mu_\mr{R}}

\newcommand{\emvvjj}{\nu_e e^+ \nu_\mu \mu^+jj}
\newcommand{\emmvjj}{\nu_e e^+\mu^-\mu^+jj}
\newcommand{\evmvjj}{e^+\nu_e\mu^-\bar\nu_{\mu} jj}
\newcommand{\eemmjj}{e^-e^+\mu^-\mu^+jj}
\newcommand{\VBFNLO}{{\textsc {Vbfnlo}}}
\newcommand{\HELAC}{{\textsc {Helac-Dipoles}}}
\newcommand{\TeV}{\unskip\,\mathrm{TeV}}

\newcommand{\evmvbb}{e^+\nu_e\mu^-\bar\nu_{\mu} b\bar{b}}
\newcommand{\evmvbbj}{e^+\nu_e\mu^-\bar\nu_{\mu} b\bar{b}j}
\newcommand{\evmvbbjj}{e^+\nu_e\mu^-\bar\nu_{\mu} b\bar{b}jj}

\newcommand{\beq}{\begin{equation}}
\newcommand{\eeq}{\end{equation}}

\begin{document}

\begin{center}
\section*{Physics Opportunities
for Vector-Boson Scattering at 
a Future  100 TeV Hadron Collider}

\vspace{2.5cm}
{\bf{B.~J\"ager$^{\,1}$, L.~Salfelder$^{\,1}$, M. Worek$^{\,2}$,
    D.~Zeppenfeld$^{\,3}$}}

\end{center}

\vskip 2.5cm

\begin{center}
{\it 
$^{1}$ Institute for Theoretical Physics, University of T\"ubingen,
Auf der Morgenstelle 14, 72076 T\"ubingen,  Germany \\ \noindent
$^{2}$ Institut for Theoretical Particle Physics and Cosmology, 
RWTH Aachen University,  52056 Aachen, Germany \\ \noindent
$^{3}$ Institute for Theoretical Physics, KIT, 76128 Karlsruhe,
Germany \\ 
}
\noindent
\end{center}
\vfill

\begin{abstract}
Vector-boson scattering (VBS) processes provide particularly promising
means for probing the mechanism of electroweak symmetry breaking and
to search for new physics in the weak sector. In the environment of a
future proton-proton collider operating at a 
center-of-mass energy of 100~TeV,
unprecedented opportunities arise for the investigation of this
important class of reactions. We highlight the prominent features of
VBS processes in this energy regime and discuss how the VBS signal can
be isolated in the presence of a priori large QCD backgrounds. We find
excellent opportunities for the analysis of VBS reactions in a
kinematic range that is inaccessible to present colliders.
\end{abstract}

\vspace{2.5cm}
TTK-17-04
\newpage

%
\section{Introduction}
%

With the discovery of the Higgs boson at the CERN Large Hadron
Collider (LHC) by the ATLAS~\cite{Aad:2012tfa} and CMS
collaborations~\cite{Chatrchyan:2012xdj} particle physics has entered
a new era. However, with the existence of the last missing building
block of the Standard Model (SM) being experimentally confirmed,
understanding the very mechanism responsible for electroweak symmetry
breaking still requires advanced efforts. Complementary to approaches
seeking to probe the form of the Higgs potential via measurements of
triple and quartic self-interactions of the Higgs boson in processes
that are plagued by low event rates and large backgrounds
\cite{Dolan:2012rv,Dolan:2015zja,Baglio:2015wcg,Baglio:2012np,Plehn:2005nk,Djouadi:1999rca,
Baur:2002qd}, weak boson scattering reactions provide promising means
to explore the electroweak sector of the SM in an experimentally more
easily accessible regime
\cite{Duncan:1985vj,Bagger:1995mk,Butterworth:2002tt}. Weak boson
scattering processes, i.e. the genuine $2\to 2$ scattering processes
$VV\to VV$ (with $V$ denoting a $W^\pm$ or a $Z$ gauge boson), are
particularly sensitive to new physics in the electroweak gauge boson
sector. For instance, in the absence of a light Higgs boson, the
amplitudes for the scattering of the longitudinal gauge boson modes,
$V_LV_L\to V_LV_L$, would grow unphysically with energy, giving rise
to unitarity violations in the TeV regime.  Alternatively, new
resonances in the weak sector, such as a $Z^\prime$ boson, could
result in prominent peaks in the invariant mass of the final-state
$VV$ system. Since so far no explicit deviations of the SM have been
observed experimentally in the weak sector, one expects new physics
effects either to be very strongly suppressed or to become visible
only in energy regimes that have not been accessible experimentally so
far.

At hadron colliders, weak vector-boson scattering (VBS) processes are
probed via the scattering of quarks or anti-quarks by the exchange of
weak bosons in the $t$-channel followed by subsequent emission of weak
bosons, i.e. purely electroweak reactions of the type $qq\to
qqVV$. The weak gauge bosons in turn decay, either leptonically or
hadronically. Because of the color-singlet nature of the weak-boson
exchange in these reactions, VBS processes exhibit particularly
distinctive features in the detector, the scattered quarks giving rise
to two tagging jets in the forward and backward
regions, while the decay products of the weak bosons are located in
the central-rapidity region. A fairly small scattering
  angle of those very energetic tagging jets is reflected in their large
  invariant mass and rather moderate transverse momenta. These
characteristics are essential in distinguishing the signal of interest
from a priori very large contributions of QCD-induced background
processes. Strategies for optimizing signal-to-background ratios for
VBS processes at the LHC have been explored in detail, resulting in
the first observation of a VBS signal in the $\wpwpjj$ mode by the
ATLAS~\cite{Aad:2014zda} and CMS
collaborations~\cite{Khachatryan:2014sta} at a center-of-mass energy
of $\sqrt{s}=8$ TeV. While only very loose exclusion bounds on new
physics effects in the weak sector could be placed with the limited
event rates accessible at 8 TeV, stronger bounds are expected from
data taken at the higher energy of 13~TeV in LHC Run~II.

Much more powerful means for probing VBS processes would be provided,
however, by a future high-energy hadron collider operating at an
energy of 100~TeV, such as a Future Circular Collider (FCC) currently
discussed as a follow-up project of the CERN LHC, or a Chinese Super
proton-proton Collider (SppC) \footnote{In the following, for
simplicity we will use the acronym FCC to generically refer to
proton-proton colliders operating at an energy of 100~TeV. }.
In order to explore the physics capabilities of such a machine, in
\cite{Mangano:2016jyj} we have performed a preliminary study of VBS
processes in the presence of background reactions at the FCC. 
A related analysis was performed in Ref.~\cite{Goncalves:2017gzy}.  
Here, we
go beyond that work and present a detailed signal-background analysis
for VBS processes taking into account  the following production modes:
$\wpwpjj$, $\wpzjj$, $\wpwmjj$, and $\zzjj$. In all cases leptonic
decays of both gauge bosons are considered. Moreover, spin
correlations of the decay products, finite width as well as off-shell
effects of gauge bosons in the signal and background processes are
fully included.

The paper is organized as follows: The general setup of our analysis
is described in Sec.~\ref{sec:setup}.  A generic way for the
parameterization of new physics effects as provided by an effective
field theory expansion is presented in Sec.~\ref{sec:eft}. In
Sec.~\ref{sec:sm} for each mode we devise a set of customized cuts to
optimize the signal in the presence of QCD-induced $VVjj$
backgrounds. In the case of $pp\to\wpwmjj$ production,
in addition to the QCD induced $pp\to\wpwmjj$ process contributions
from the overwhelming $t\bar t$ background processes are considered.
Here additional techniques are applied to suppress this background in
order to improve the signal-to-background ratio. Our conclusions are
presented in Sec.~\ref{sec:summ}.

%
\section{Setup of the analysis}
\label{sec:setup}
%

Throughout this analysis, for each $pp\to VVjj$ VBS process and its
associated QCD backgrounds we focus on a particular decay mode with a
clean experimental signature.  We assume that each weak
boson $V$ decays leptonically and we consider decays of the $VV$ systems only into
two different generations of leptons. For instance, for the VBS-induced
$\wpwpjj$ production mode we consider the $\emvvjj$ final state at
${\cal O}(\alpha^6)$. Other lepton combinations in the final state
(such as $\nu_e e^+ \nu_e e^+$ and $\nu_\mu \mu^+
  \nu_\mu \mu^+$)  can be obtained thereof straightforwardly, as long
as same-type lepton interference effects are neglected.
These interference effects are completely negligible, in
particular in the kinematic range we are interested in, with large
invariant masses of the four-lepton system that disfavor same-type
lepton interference configurations.  
The cross section summed over different lepton flavors ($\ell =
e,\mu$) can be obtained by multiplying the result for a single
leptonic flavor channel with a suitable combinatorial factor.
Off-shell and non-resonant contributions of the gauge bosons are fully
taken into account.  For simplicity, we nonetheless refer to the
respective process as ``$VVjj$ production'', implicitly assuming that
decays of the weak bosons are taken into account. Identical-flavor
interference effects between $t$- and $u$-channel diagrams are
neglected for all signal processes. 
Triple gauge-boson production processes of the type
$pp\to VVV$ with subsequent decay of a weak boson into a pair of jets
are assumed to be part of a different reaction and not considered
here.  Even though such processes in principle give rise to the same
final state as VBS processes, they exhibit entirely different
kinematic properties and can be safely neglected after the application
of VBS specific selection cuts.
The Cabibbo-Kobayashi-Maskawa (CKM) mixing of the quark generations is
neglected, i.e.\ a diagonal form is assumed for the CKM matrix, since
mixing within the first two generations cancels exactly within the
$t$-channel approximation as no charm tagging is attempted.
For all $VVjj$ processes, contributions from external top or bottom
quarks are regarded as separate processes not to be considered here.

For the QCD background processes all Feynman diagrams that result in
the specified final state at ${\cal O}(\alpha_s^2\alpha^4)$ are
accounted for with all interference effects as well as non-zero gauge
boson width effects.  We shall refer to these processes as ``$VVjj$
QCD production".
In the same way we proceed for background processes with top-quarks.
We consider $t\bar{t}$, $t\bar{t}j$ and $t\bar{t}jj$ production,
respectively, at ${\cal O}(\alpha_s^2\alpha^4)$, ${\cal
O}(\alpha_s^3\alpha^4)$, ${\cal
O}(\alpha_s^4\alpha^4)$. We include all double-,
single- and non-resonant diagrams, interference effects as well as
off-shell effects of the top quarks. Additionally, non-resonant and off-shell
effects due to the finite $W$-boson width are taken into account.  We
always combine these three background processes into the
``$t\bar{t}+\mr{jets}$" sample.  We note here that for
  all top-quark processes massive final-state bottom quarks will be considered.  
Calculations for all signal and background processes are performed at
 the leading order (LO). 

The numerical simulations that form the basis of our discussion below
make use of the \VBFNLO{} package \cite{Arnold:2008rz} for the VBS
signal processes. All background processes are computed with the
\HELAC{} package \cite{Czakon:2009ss} and cross-checked with the
\textsc{Helac-Phegas} Monte-Carlo program
\cite{Cafarella:2007pc}. Both the \textsc{Helac-Dipoles} and
\textsc{Helac-Phegas} programs use for the calculation of scattering
amplitudes an automatic off-shell iterative algorithm
\cite{Kanaki:2000ey,Papadopoulos:2005ky,Bevilacqua:2011xh}.
For the phase-space integration, \textsc{Phegas}
\cite{Papadopoulos:2000tt} in conjunction with \textsc{Parni}
\cite{vanHameren:2007pt} as well as \textsc{Kaleu} \cite{vanHameren:2010gg}
have been employed.

In order to warrant the correctness of our results we
have cross-checked \HELAC{} results with \VBFNLO{} for all processes
that are implemented in the \VBFNLO{} program. 
  In each case we found
full agreement for cross sections and distributions within the
numerical accuracy of the two codes.

Throughout, for the masses $m$ and widths $\Gamma$ of the $W$, $Z$,
and $H$ bosons we use the following values:
\begin{equation}
\begin{array}{cclclll}
m_W&=&80.385~\mathrm{GeV}, &\quad \quad \quad &  
\Gamma_W &=& 2.097547~\mathrm{GeV}\,,
\vspace{0.2cm}\\
m_Z &=&91.1876~\mathrm{GeV}, &\quad \quad \quad & 
\Gamma_Z &=& 2.508827~\mathrm{GeV}\,,
\vspace{0.2cm}\\
m_H &=&125.09~\mathrm{GeV}, & \quad \quad \quad & 
\Gamma_H &=& 0.004066~\mathrm{GeV}\,. 
\end{array}
\end{equation}
Further electroweak (EW) parameters such as the EW coupling $\alpha$
and the weak mixing angle $\sin\theta_W$ are computed in the $G_\mu$
scheme with the Fermi constant $G_{F}=1.1663787\times10^{-5}~{\rm
GeV}^{-2}$.
The mass and width of the top quark are set to 
\begin{equation}
\begin{array}{cclclll}
m_{top}&=&172.5~\mathrm{GeV},&\quad \quad \quad  & 
\Gamma_{top} &=&
 1.4576~\mathrm{GeV}\,. 
\end{array}
\end{equation}
For the parton distribution functions (PDFs) of the
proton, we use the MMHT2014lo68cl set~\cite{Harland-Lang:2014zoa} for
all electroweak and QCD-induced $VVjj$ processes.  For
$t\bar{t}+\mr{jets}$ production  the
MMHT2014lo68cl${}_{-}$nf4  PDF set is used instead.  Consequently, 
in the former case we use the
so called five-flavor-scheme, whereas, in the latter we employed the
so-called  four-flavor scheme with only gluons and light-flavor
quarks in the proton, where massive bottom quarks are produced from
gluon splitting at short distances.  The mass of the bottom quark is
set to 
\begin{equation}
m_b=4.75 ~{\rm GeV}\,.
\end{equation}
The associated strong coupling $\alpha_s$
running at one loop is provided by the LHAPDF repository
\cite{Buckley:2014ana}. 

The factorization and renormalization scales, $\muf$ and $\mur$, are
set process-specifically: The VBS signal processes $pp\to VVjj$ with
fully leptonic decays are of the order of $\mc{O}(\alpha^6)$.
The leading order VBS results are, thus, independent of
$\mu_R$.  For $\mu_F$, we use the momentum transfer $Q_i$ of
the incoming to the outgoing (anti-)quark on the upper and lower
fermion lines, respectively,
\beq
\mu_F= Q_i\,.  
\eeq
For the QCD-induced $VVjj$ production processes, which are of the
order of ${\cal O}(\alpha_s^2\alpha^4)$, the total
transverse energy $H_T^{VV}$ of each event is used as reference scale,
such that
\beq
\mu_F=\mu_R=H_T^{VV}/2\,,
\eeq
with 
\begin{equation}
H_T^{VV} = \sum_{i} p_{T,\,i}+E_T(V_1)+E_T(V_2)\,. 
\end{equation} 
Here, the summation runs over the transverse momenta 
$p_{T,\,i}$ of all final-state partons involved in an event, while
the transverse energy of each gauge boson is computed from its mass
and the momenta of its leptonic decay products according to
\begin{equation}
E_T(V_i)=\sqrt{p_{T}^2(V_i)+m_{V_i}^2}\,,
\end{equation}
where $i$ denotes a $W^\pm$ or $Z$~boson. Similarly, for the
top-quark induced background processes, that are respectively of the
order of  ${\cal O}(\alpha_s^2\alpha^4),\, {\cal
    O}(\alpha_s^3\alpha^4)$  and ${\cal O}(\alpha_s^4\alpha^4)$, we use
\beq
\mu_F=\mu_R=H_T^\mr{top}/2\,,
\eeq
with the transverse energy $H_T^\mr{top}$ being computed from the
transverse energy of the top quarks or anti-quarks,
\begin{equation}
E_T(t)=\sqrt{p_{T}^2(t)+m_{top}^2}\,, 
\quad \quad \quad \quad \quad \quad
E_T(\,\bar{t}\,)=\sqrt{p_{T}^2(\,\bar{t}\,)+m_{top}^2}\,.
\end{equation}
and the transverse momenta of all light partons in the final state of
an event, 
\begin{equation}
H_T^\mr{top}= \sum_{i} p_{T,\,i}+E_T(t)+E_T(\,\bar{t}\,)\,.
\end{equation}
In the following sections we will derive dedicated selection cuts for
each VBS process. However, a common set of minimal
selection cuts will be applied in each case: In order to
reconstruct jets from partons, we use the anti-$k_T$  jet
algorithm \cite{Cacciari:2008gp} with $R=0.4$.  
We require at least two jets. The two hardest jets of each event 
are called ``tagging jets'' and need to have  a minimum transverse
momentum of 
\begin{equation}
\label{eq:jet-def}
p_{T,\,\mr{jet}} \ge 50~\mathrm{GeV}\,.
\end{equation}
The two tagging jets are  also required to reside in
  opposite hemispheres of  the detector, 
\begin{equation}
\label{eq:ysign}
y_{j_1}^\mr{tag}\times y_{j_2}^\mr{tag} < 0\,. 
\end{equation}
Charged leptons need to fulfill cuts on transverse momenta, rapidities, and be well separated from any jet in the rapidity-azimuthal angle plane,
\begin{equation}
\label{eq:ptl-cut}
p_{T,\,\ell} \ge 20~\mathrm{GeV}\,,\qquad
|y_{\ell}| \le 5 \,,\qquad  
\Delta R_{\mr{jet},\,\ell} \ge 0.4\,.
\end{equation}
A very powerful tool for the suppression of background
processes is provided by requiring that all charged leptons are
located between the two tagging jets in rapidity
\begin{equation}
\label{eq:lgap}
 y_{j,\,min}^\mr{tag} < y_{\ell}  < y_{j,\,max}^\mr{tag}\,. 
\end{equation}
To suppress contributions from virtual photons,
$\gamma^*\to \ell^+\ell^-$, we furthermore impose the minimal
invariant-mass cut on oppositely charged (same flavor) 
leptons, that occur in  both $ZZjj$ and $W^\pm Z jj$ processes. Thus,
in these channels we require 
\begin{equation}
\label{eq:ll-cuts}
M_{e^+e^-} > 15~\mathrm{GeV}\,,
\quad 
M_{\mu^+\mu^-} > 15~\mathrm{GeV}
\,.  
\end{equation}

%
\section{Effective field theory expansion}
\label{sec:eft}
%

A generic way for the parameterization of new physics effects is
provided by an effective field theory (EFT) expansion. Such an
effective theory can be constructed as a low-energy approximation of a
more fundamental theory, and it is valid only up to a specific energy
scale $\Lambda$.  The Lagrangian of an EFT is typically expressed in
terms of the SM Lagrangian and additional terms including operators
$\mc{O}_i^{(d)}$ of higher dimensionality $d$,
\beq
\label{eq:eft}
\mc{L}_\mr{EFT} = \mc{L}_\mr{SM} + \sum_{d>4}\sum_i
\frac{f_i^{(d)}}{\Lambda^{d-4}}\mc{O}_i^{(d)}\,,
\eeq
with $f_i^{(d)}$
denoting the coefficients of the expansion. As
long as an energy regime far below the breakdown scale $\Lambda$ of
the EFT expansion is considered, the non-SM part of the EFT expansion
is dominated by the lowest relevant terms, i.e.\ contributions due to
operators of dimension six and eight. 

Since dimension-six operators affect quartic as well as triple
gauge boson couplings, they can most conveniently be probed in
gauge-boson pair production processes that exhibit the respective
triple-gauge boson vertices. The dimension-eight operators below, instead, do
not have any impact on triple but only on quartic gauge boson
couplings that emerge in triple-vector boson production and VBS
reactions. The experimentally clean VBS processes thus represent a
particularly important test bed for this class of new-physics
contributions (see, e.g.\ Ref.~\cite{Rauch:2016pai} for a recent review).
In the present study, we therefore consider the impact of new physics
on VBS processes in the EFT approach that is due to CP conserving
operators of dimension eight that modify quartic weak gauge boson
couplings.  Following the notation of
Refs.~\cite{Hagiwara:1993ck,Degrande:2013ng,Eboli:2006wa}, we consider 
a set of representative dimension-eight operators, $\mc{O}_i^{(d=8)}$, which in the literature are  referred to as $\mc{O}_{S,1}$, $\mc{O}_{M,1}$, and $\mc{O}_{T,0}$, respectively. They are defined as 
\begin{align}
\label{eq:CP-con8}
\mc{O}_{S,1}&=\left[(D_\mu \Phi)^\dagger (D^\mu \Phi)\right]\times  
					\left[(D_\nu \Phi)^\dagger (D^\nu \Phi)\right]\,, \\
\mc{O}_{M,1}&=\mr{Tr}\left[\hat{W}_{\mu\nu}\hat{W}^{\nu\beta}\right]\times  
\left[(D_\beta \Phi)^\dagger (D^\mu \Phi)\right]\,, \\
\mc{O}_{T,0}&=\mr{Tr}\left[\hat{W}_{\mu\nu}\hat{W}^{\mu\nu}\right]\times  
\mr{Tr}\left[\hat{W}_{\alpha\beta}\hat{W}^{\alpha\beta}\right] \,, 
\end{align}
with the associated coefficients $f_{S,1}/\Lambda^4$, $f_{M,1}/\Lambda^4$,
and $f_{T,0}/\Lambda^4$.
To simplify the notation we drop the dimensionality index $(d=8)$ here and
in the  following.
The considered operators contain the Higgs-doublet field $\Phi$, the covariant
derivative $D_\mu$, and the field-strength tensor $W_{\mu\nu}$, defined as
\begin{align}
D_\mu&=\partial_\mu+\frac{i}{2}g\tau^I W_\mu^I+\frac{i}{2}g'B_\mu\,, \\
W_{\mu\nu}&=\frac{i}{2}g\tau^I(\partial_\mu W^I_\nu-\partial_\nu
            W^I_\mu -g\epsilon_{IJK}W^J_\mu W^K_\nu)\,, 
\end{align}
with the U(1) and SU(2) gauge fields $B_\mu$ and $W_\mu$, the
associated couplings $g^\prime$ and $g$, the weak isospin matrices
$\tau^I$, and  
\beq
\hat{W}_{\mu\nu} = i\frac{g}{2}W_{\mu\nu}^a \sigma^a\,.
\eeq

By construction, the EFT approach is valid only in a
restricted energy regime. If the EFT
expansion is truncated, unphysical violations of unitarity may occur
beyond some scale $\Lambda_{U}$. Such unitarity violations can be
avoided, if the EFT is restricted to the region where it is fully
valid.  Practically, this can be achieved by form factors that
suppress EFT contributions above scales where the EFT expansion is
supposed to loose its applicability.  In the code package \VBFNLO{}
the impact of dimension-six and dimension-eight operators on
VBS processes is explicitly accounted for. The user can set the relevant
operator coefficients to customized values.  In order to maintain
unitarity, a form factor 
\beq
\label{eq:ff}
F =
	\left\{
	\begin{array}{ccc}
		1  & \ldots & M_{VV}< \Lambda_F
		\\
		\left(\frac{\Lambda_F}{M_{VV}}\right)^4 & \ldots & M_{VV}> \Lambda_F
	\end{array}
	\right.	
\eeq
can be applied, where $M_{VV}$ denotes the invariant mass of the produced
$V$~bosons in $VVjj$ reactions. Note that $\Lambda_{F}\leq\Lambda_{U}$, and $\Lambda_{F}$ is related to $\Lambda$, but not necessarily identical. 

%
\section{Signal and background processes}
\label{sec:sm}
%

In order to fully exploit the capabilities of a future
100~TeV proton-proton collider for the analysis of VBS processes,
optimized selection cuts for each specific final state need to be
devised.  In the following, we present process-specific selection
criteria for various VBS channels and discuss the behavior of signal
and background contributions for characteristic kinematic
distributions.

%
\subsection{$\wpwpjj$}
\label{sec:wpwpjj}
%

\begin{figure}
\center
\includegraphics[angle=0,scale=0.7,bb=50 380 330
715]{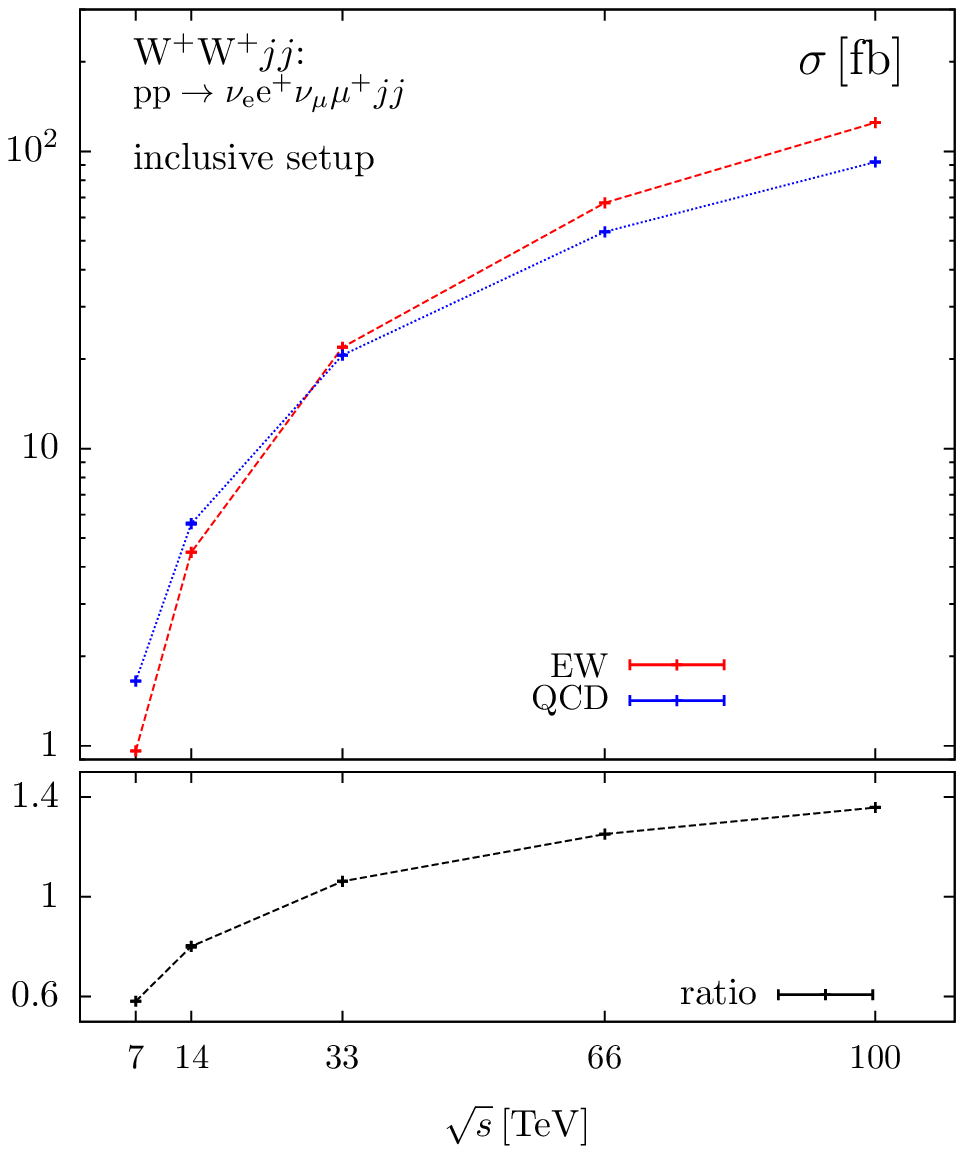}
\caption{\it Energy dependence of the EW-induced (red line) and
QCD-induced (blue line) contributions to the inclusive cross section
for $pp\to \nu_e e^+\nu_\mu \mu^+ jj$, without any selection cuts. The lower
panel shows the ratio of the EW to the QCD contribution.
\label{fig:wpwpjj-xsec}
}
\end{figure}

The same-sign diboson-plus-dijet final state, $pp\to \wpwpjj$,
provides a particularly clean signature in the fully leptonic decay
mode, as the only irreducible background in the $\emvvjj$ channel
comprises the QCD-induced $\wpwpjj$ production process. In
contrast to other VBS channels that receive large background
contributions from gluon-induced QCD processes, same-sign diboson
production can only proceed via the scattering of (anti-)quarks of
proper type to ensure the conservation of electromagnetic charge
throughout the reaction.  In the next-to-leading-order QCD analysis of
\cite{Jager:2011ms}  it was shown that with VBS-specific selection
cuts a signal-to-background ($S/B$) ratio  of about 27 could be
achieved at the LHC operating at an energy of 7~TeV. This ratio seems
remarkable at first sight, if one naively assumes the size of the
respective EW and QCD cross sections to be determined by the relevant
orders of the strong and electromagnetic couplings that themselves differ by roughly one order of
magnitude. Upon closer inspection it becomes clear that the {\em
inclusive} cross section for the QCD-induced $\emvvjj$ final state
indeed exceeds the EW one by about a factor of 1.7. However, the
very distinct kinematic properties of the VBS production mode allow
the efficient suppression of the QCD background contributions with a
dedicated set of selection cuts that diminish the rate of signal
events only marginally, resulting in the large $S/B$ ratio reported
above.  At higher collider energies, the production rate for the EW
signal process increases slightly faster than the production rate for
the QCD background, as depicted in Fig.~\ref{fig:wpwpjj-xsec}. 
%
%
We will see below that an approach similar to the strategy
applied at the 7~TeV LHC can be used for higher center-of-mass 
energies yielding even better $S/B$ ratios at a 100~TeV collider.

\begin{figure}
\includegraphics[angle=0,width=0.5\textwidth]{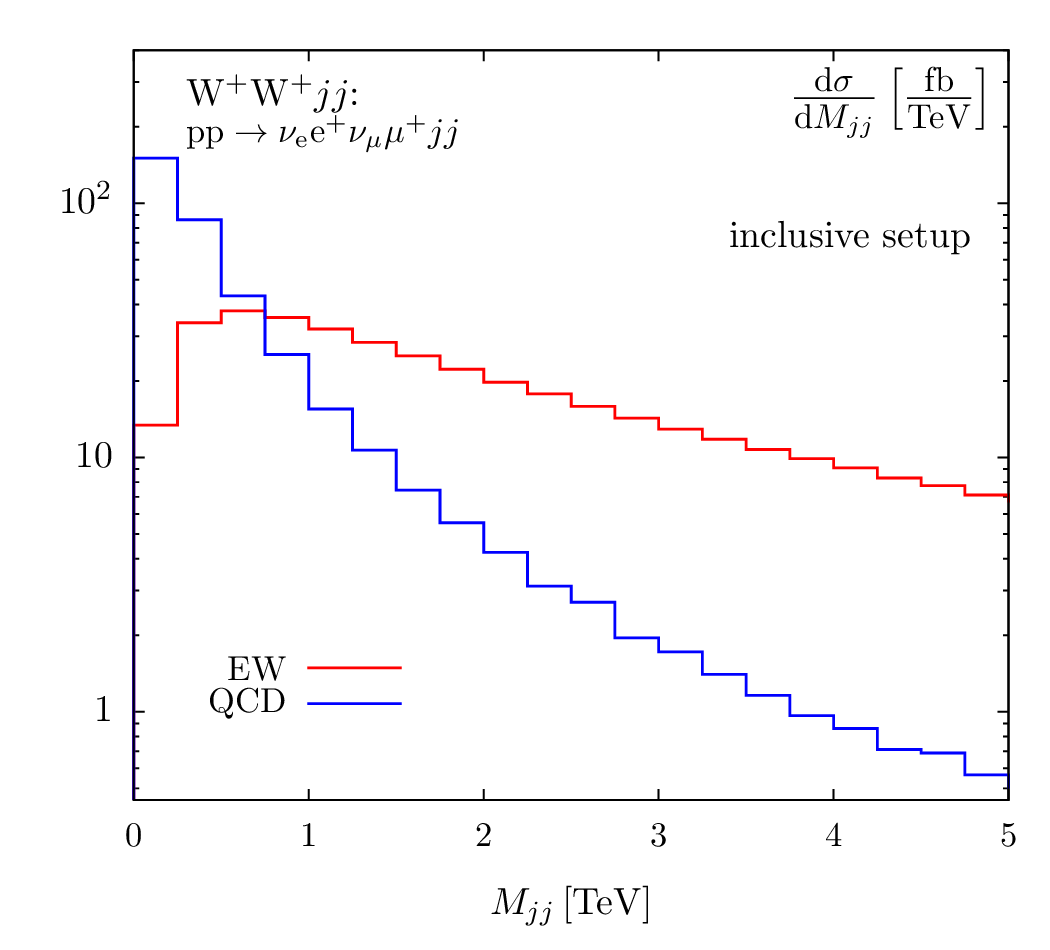}
\includegraphics[angle=0,width=0.5\textwidth]{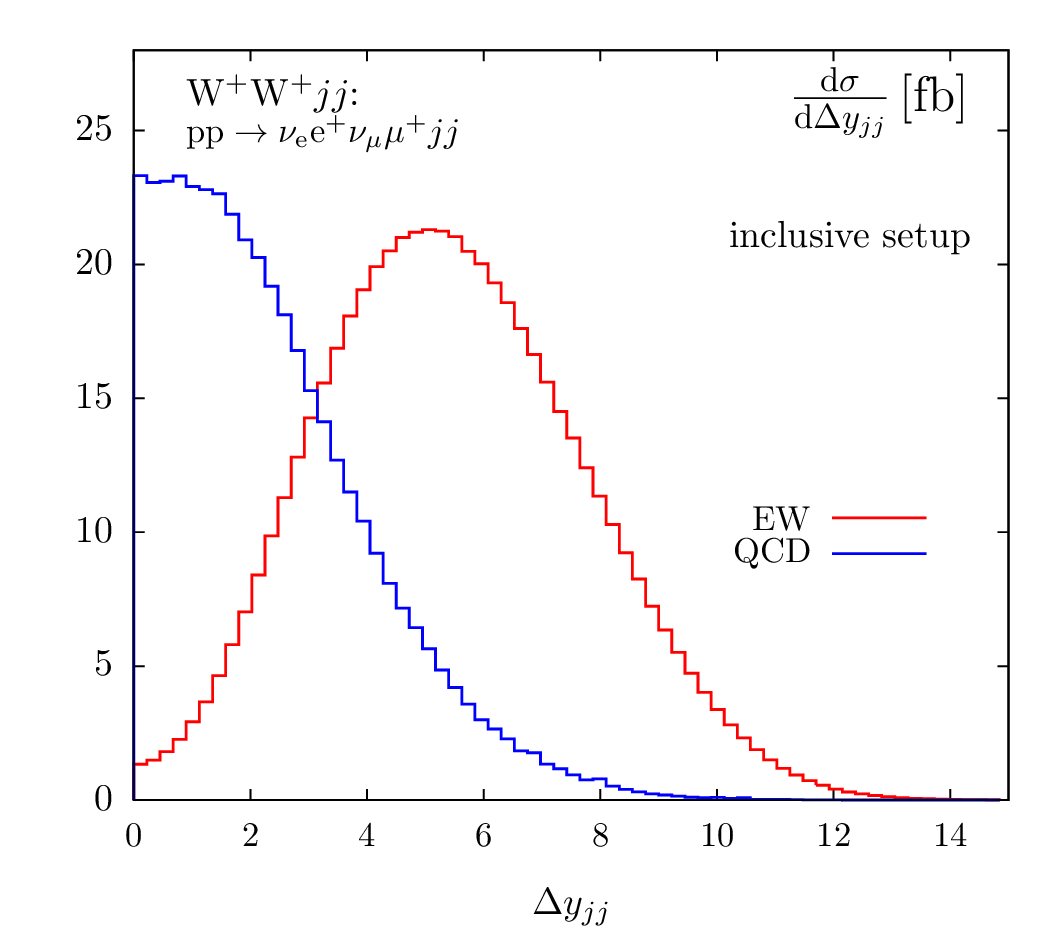}
\caption{\it Invariant mass~(l.h.s.) and rapidity separation of the two
tagging jets~(r.h.s) for the EW-induced (red lines) and QCD-induced
(blue lines) contributions to $pp\to\nu_e e^+\nu_\mu \mu^+ jj$,  without
any selection cuts. 
\label{fig:wpwpjj-inc}
}
\end{figure}

\begin{figure}
\includegraphics[angle=0,width=0.5\textwidth]{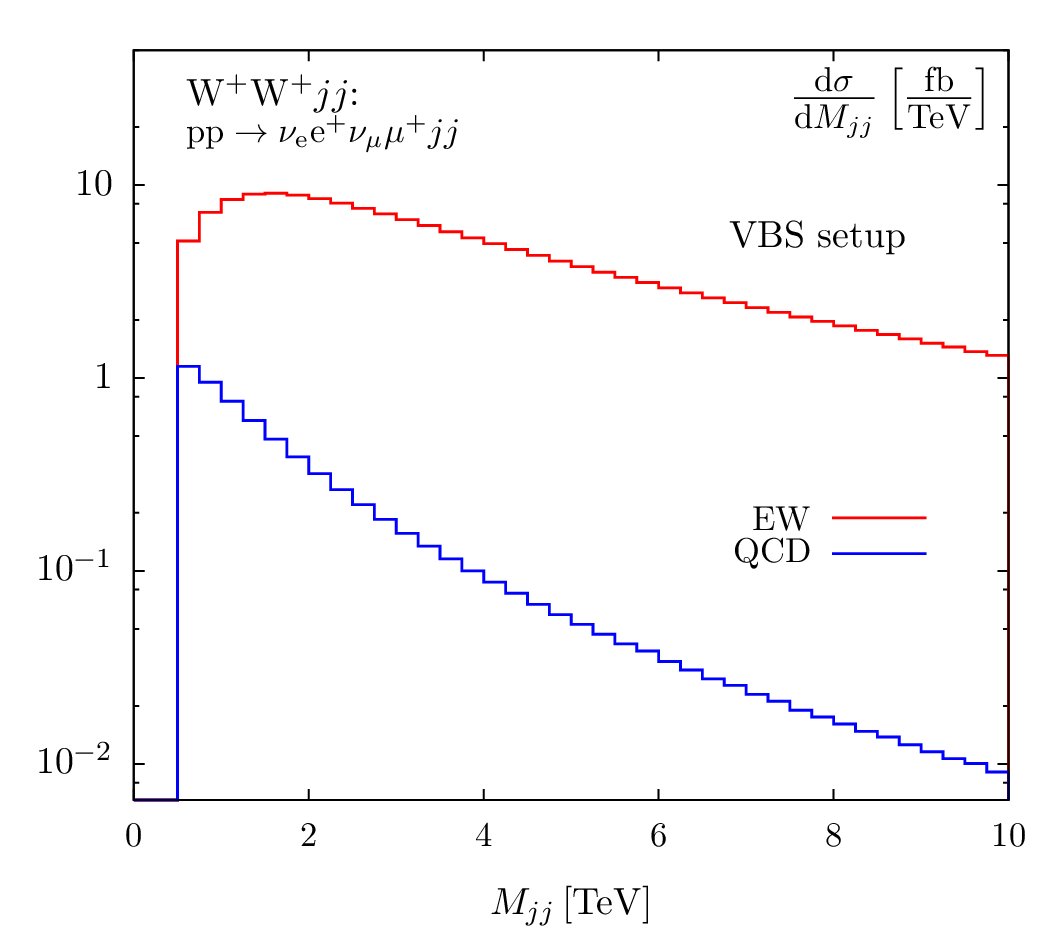}
\includegraphics[angle=0,width=0.5\textwidth]{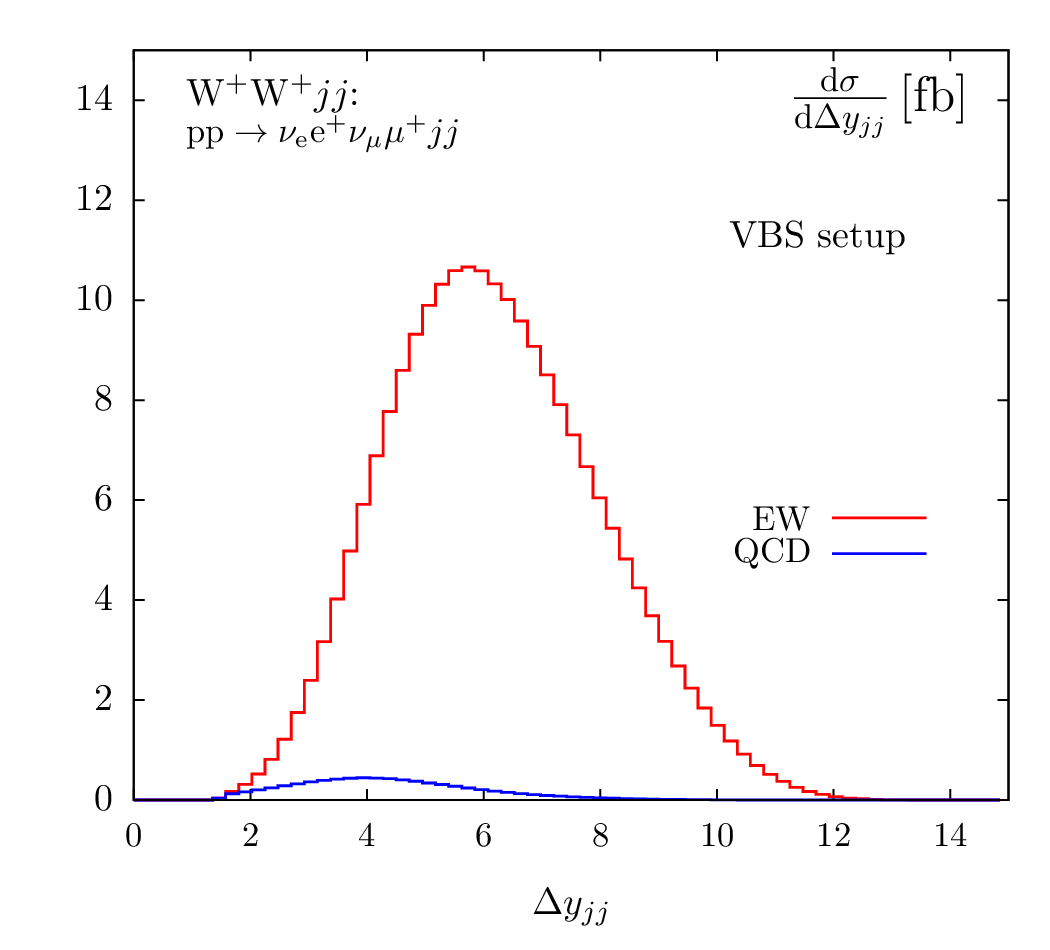}
\caption{\it Invariant mass~(l.h.s.) and rapidity separation of the two
tagging jets~(r.h.s) for the EW-induced (red lines) and QCD-induced
(blue lines) contributions to $pp\to \nu_e e^+\nu_\mu \mu^+ jj$,
within the selection cuts of Eqs.~(\ref{eq:jet-def})--(\ref{eq:lgap})
and Eq.~(\ref{eq:jjcuts_wpp}).
\label{fig:wpwpjj-cuts}
}
\end{figure}

To illustrate the capability of selection cuts in the environment of
an FCC, focusing on a fully inclusive setup (i.e. not imposing any
selection cuts) in Fig.~\ref{fig:wpwpjj-inc} we present two
distributions that exhibit particularly distinctive shapes in
VBS-induced processes: the invariant mass $M_{jj}$ of the tagging
jets' system and the rapidity separation of the two tagging jets,
$\Delta y_{jj}$.
%
%
While in QCD-induced $\wpwpjj$ processes the two jets are
mostly produced with a small invariant mass and close to each other in
rapidity, the color-singlet nature of the weak-boson exchange, which
is characteristic for VBS processes, gives rise to a dijet system of
large invariant mass with a substantial separation in rapidity.  
This feature can be exploited for the design of powerful
selection cuts. It can be easily seen from
Fig.~\ref{fig:wpwpjj-inc} that the $\Delta y_{jj}$ distribution (right
panel) peaks around zero values for the QCD background, while the VBS
contribution exhibits a dip there, and peaks around $\Delta y_{jj}
=5$. On the other hand, a steep rise of the invariant mass
distribution towards small values of $M_{jj}$ can be observed in case
of the QCD induced $W^+W^+jj$ production process, where the VBS
contribution is no longer dominant.  Suitable cuts on $M_{jj}$ and
$\Delta y_{jj}$ diminish the signal cross section only
marginally, at the same time  removing  a large fraction of the QCD-induced
background contribution. Indeed, by imposing the following
process-specific selection cuts, 
\begin{equation}
\label{eq:jjcuts_wpp}
M_{jj}>500~ {\rm GeV}\,,\quad \quad \quad \quad
\Delta y_{jj}= |y_{j_1} -y_{j_2}| >1.5\,, 
\end{equation}
in addition to the generic cuts of
Eqs.~(\ref{eq:jet-def})--(\ref{eq:lgap}) we obtain cross sections of 
$\sigma_S=49.34$~fb for the signal process and 
$\sigma_B= 1.68$~fb for the
  background process,  resulting in  the $S/B$ ratio of  about 30, 
  similar to what was reported in \cite{Jager:2011ms} for the case of the LHC
operating at 7~TeV.  Figure~\ref{fig:wpwpjj-cuts} shows the invariant 
mass and rapidity separation of the tagging jet system within the
selection cuts of Eqs.~(\ref{eq:jet-def})--(\ref{eq:lgap}) and
Eq.~(\ref{eq:jjcuts_wpp}).
%

\begin{figure}
\includegraphics[angle=0,width=0.5\textwidth]{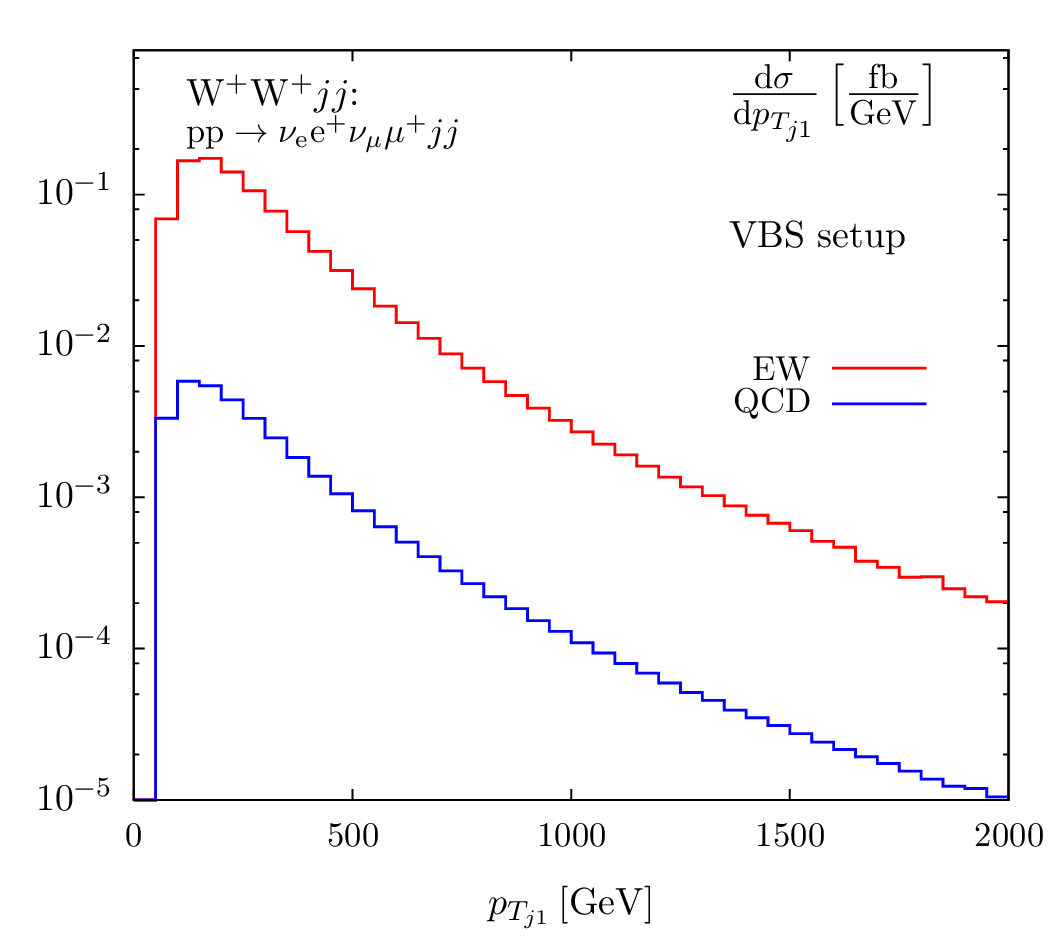}
\includegraphics[angle=0,width=0.5\textwidth]{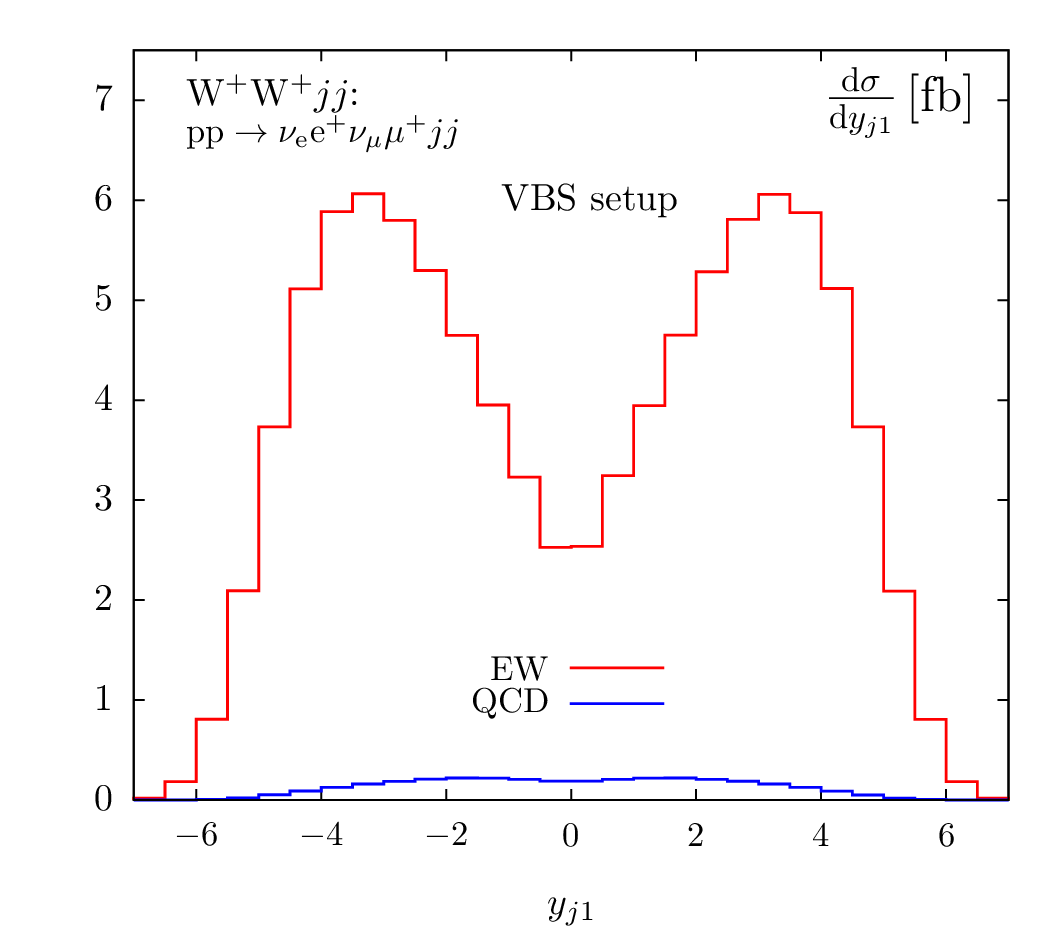}
\caption{\it Transverse-momentum~(l.h.s.) and rapidity distribution of the
hardest tagging jet~(r.h.s) for the EW-induced (red lines) and
QCD-induced (blue lines) contributions to $pp\to \nu_e e^+\nu_\mu
\mu^+ jj$, within the selection cuts of
Eqs.~(\ref{eq:jet-def})--(\ref{eq:lgap}) and
Eq.~(\ref{eq:jjcuts_wpp}).
\label{fig:wpwpjj-jets}
}
\end{figure}

\begin{figure}
\includegraphics[angle=0,width=0.5\textwidth]{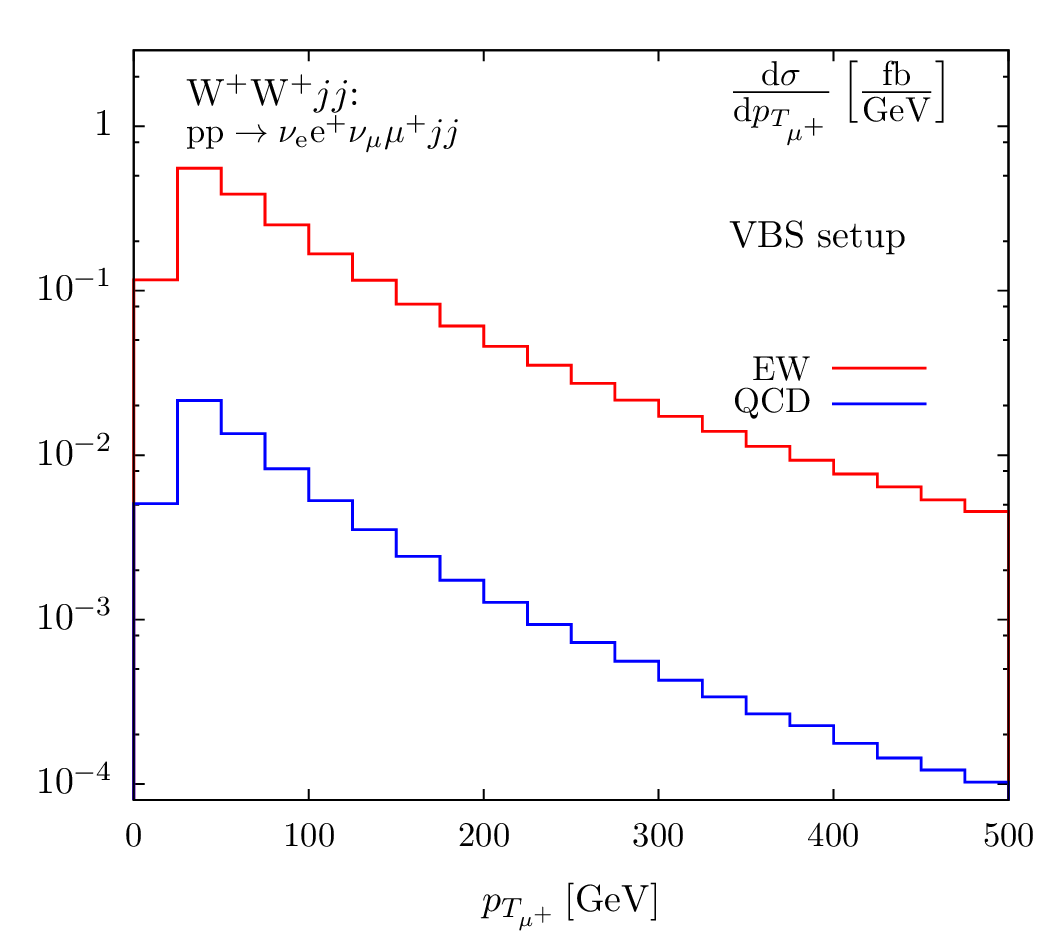}
\includegraphics[angle=0,width=0.5\textwidth]{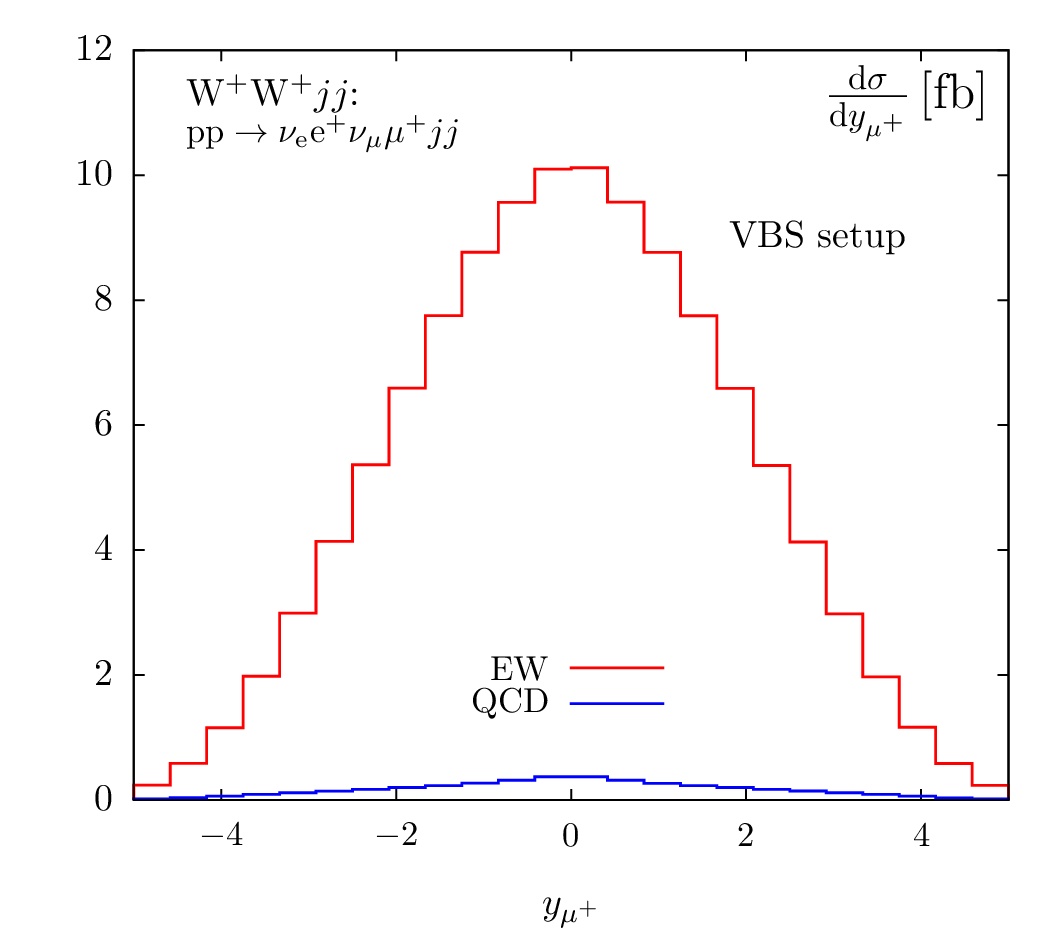}
\caption{\it Transverse-momentum~(l.h.s.) and rapidity distribution of the
muon~(r.h.s) for the EW-induced (red lines) and QCD-induced (blue 
lines) contributions to $pp\to \nu_e e^+\nu_\mu \mu^+ jj$, within the
selection cuts of Eqs.~(\ref{eq:jet-def})--(\ref{eq:lgap}) and
Eq.~(\ref{eq:jjcuts_wpp}).
\label{fig:wpwpjj-muon}
}
\end{figure}

New physics in the weak sector is expected to affect not only the shape of
distributions related to the weak bosons in $VVjj$ processes, but also
differential distributions of the tagging jets that are the
tell-tale signature of VBS reactions at hadron colliders. 
 In particular the tails of invariant-mass and
transverse-momentum distributions of final-state leptons and tagging jets
are sensitive to physics beyond the SM (BSM). At the FCC, such observables
are accessible up to much higher scales than at the LHC. Our study in~\cite{Mangano:2016jyj} revealed, for instance, that even at scales
far above 1~TeV, 
several signal events are to be expected for an integrated luminosity of 30~fb$^{-1}$.
In the $\wpwpjj$ channel, after VBS-specific selection cuts are
applied the QCD background contributions amount to only about 3\% of
the EW signal and thus have little impact on the relevant
distributions, as we demonstrate explicitly for selected observables:
Figure~\ref{fig:wpwpjj-jets} shows the transverse-momentum and the
rapidity distributions of the hardest tagging jet,
%
%
while Fig.~\ref{fig:wpwpjj-muon} illustrates the respective
distributions of the muon. 
%
%
The transverse-momentum distribution of the hardest tagging jet in the
EW signal process exhibits a peak at about 170~GeV, while the QCD
contribution tends to produce slightly softer jets. The tagging jets
 produced in the QCD mode typically are located at smaller rapidities 
than in the EW signal process where peaks in $d\sigma/dy_{j_1}$ occur
at values as large as $\pm 4$.
The muon distributions are less distinctive, as the leptons are mostly
located at central rapidities in both production modes.

\begin{figure}
\includegraphics[angle=0,width=0.5\textwidth]{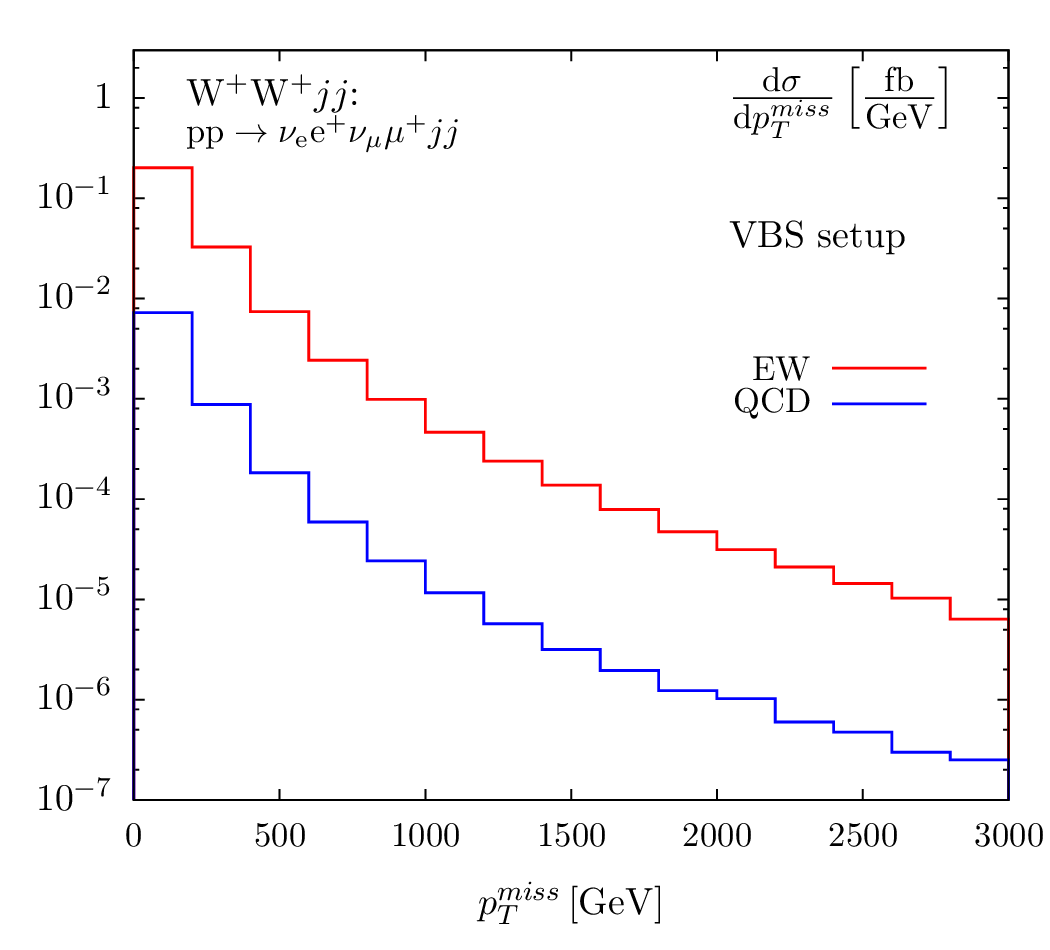}
\includegraphics[angle=0,width=0.5\textwidth]{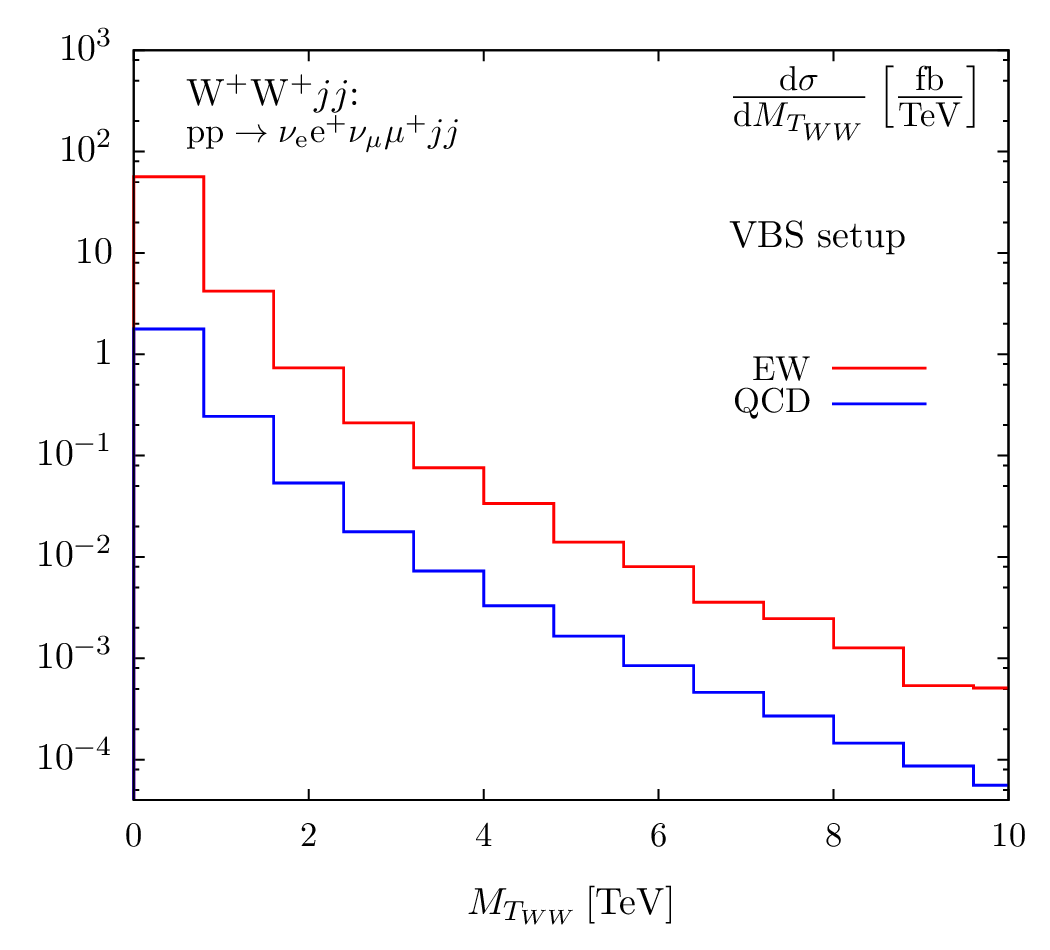}
\caption{\it Missing transverse momentum~(l.h.s.) and transverse-mass
distribution of the gauge-boson system~(r.h.s) for the EW-induced
(red lines) and QCD-induced (blue lines) contributions to $pp\to \nu_e
e^+\nu_\mu \mu^+ jj$, within the selection cuts of
Eqs.~(\ref{eq:jet-def})--(\ref{eq:lgap}) and
Eq.~(\ref{eq:jjcuts_wpp}).
\label{fig:wpwpjj-mtww}
}
\end{figure}

For narrowing down new physics searches in the weak sector, it may be
useful to consider a particular mass range of the invariant diboson
system. In the $\emvvjj$ channel, however, the invariant mass of the
$W^+W^+$ system is not fully reconstructible experimentally, because
of the presence of two neutrinos.  In this case, the transverse mass
$M_{T_{WW}}$ of the final state system consisting of two charged
leptons and missing transverse momentum in the detector, can be
considered instead. It is defined as
\begin{equation}
\label{eq:def-mtww}
M_{T_{WW}} =
 \sqrt{
           \left(E_{T}^{\,\ell\ell}+E_{T}^{\, miss}\right)^2
          -\left(\vec{p}_T^{\,\,\ell\ell}+\vec{p}_T^{\, \,miss}\right)^2
          }\,,
\end{equation}
with 
\begin{equation}
E_{T}^{\,\ell\ell} = \sqrt{(\vec{p}_T^{\,\, \ell\ell})^2+M_{\ell\ell}^2}\,,\quad 
E_{T}^{\, miss} = |\vec{p}_T^{\,\,miss}|\,.
\end{equation}
where, $\vec{p}_T^{\,\, \ell\ell}$ denotes the transverse momentum of the
charged-lepton system, and $\vec{p}_T^{\,\, miss}$  is the total
transverse momentum of the neutrino system.
The missing transverse momentum is shown together with the
transverse-mass distribution of the lepton-neutrino system for $pp\to
\nu_e e^+\nu_\mu \mu^+ jj$ in Fig.~\ref{fig:wpwpjj-mtww} in the
context of the SM.
%
%
The latter distribution is particularly interesting to study the
applicability of an EFT approach for estimating the impact of new
physics in the weak sector.  

\begin{figure}
\includegraphics[angle=0,width=0.5\textwidth]{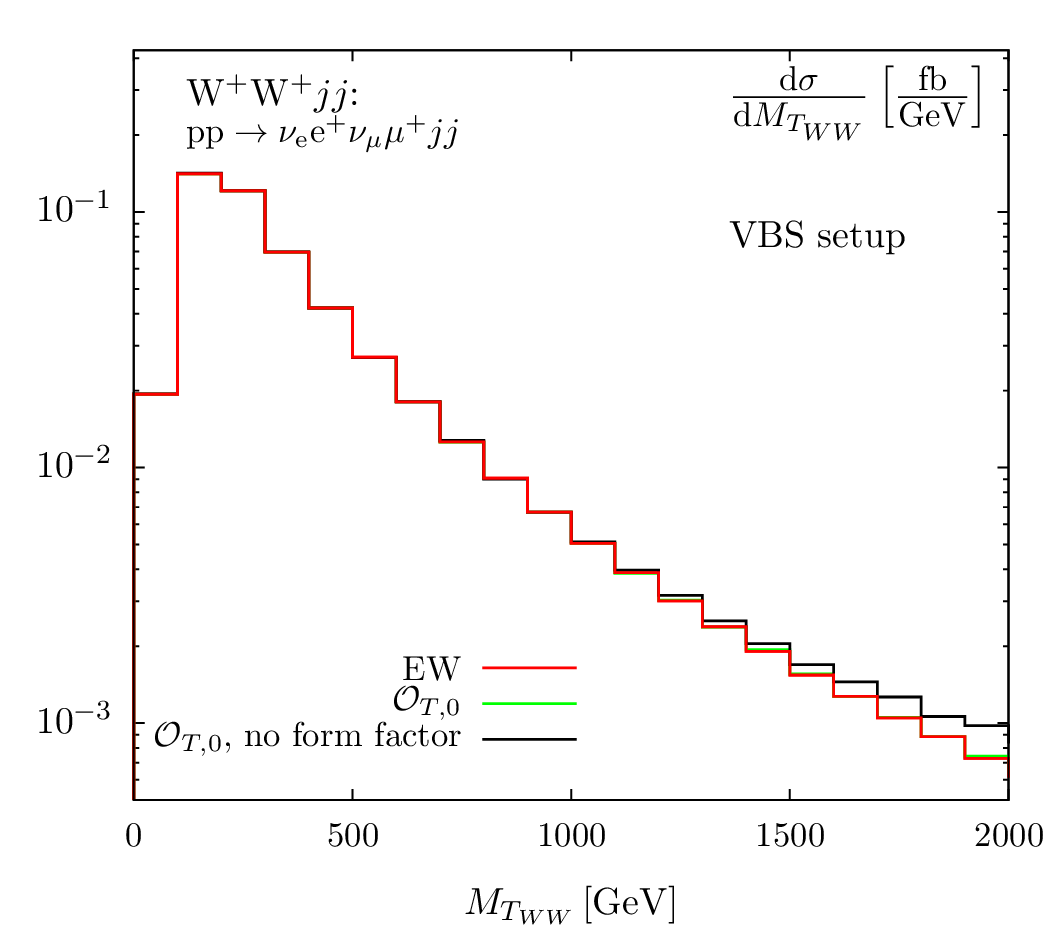}
\includegraphics[angle=0,width=0.5\textwidth]{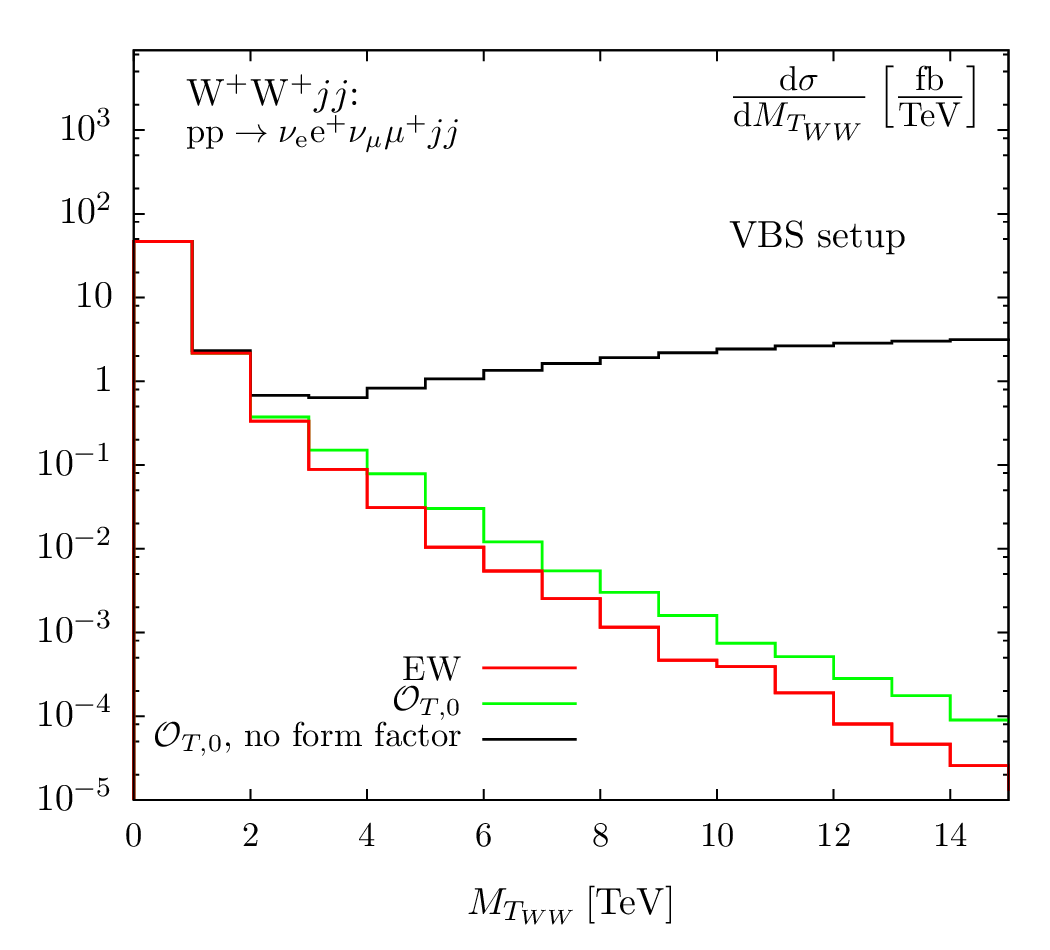}
\caption{\it 
Transverse-mass distribution of the gauge-boson system for
the process $pp\to \nu_e e^+\nu_\mu \mu^+ jj$ in the framework of the
SM (red lines) and including the impact of the dimension-eight operator 
$\mc{O}_{T,0}$ without (black lines) and with (green lines) the form
factor of Eq.~(\ref{eq:ff}), in two different plot ranges. The selection cuts of
Eqs.~(\ref{eq:jet-def})--(\ref{eq:lgap}) and Eq.~(\ref{eq:jjcuts_wpp})
are imposed.
\label{fig:wpwpjj-mtww-agc}
}
\end{figure}

Figure~\ref{fig:wpwpjj-mtww-agc}
%
%
shows $d\sigma/dM_{T_{WW}}$ for the EW signal process, both in the
context of the SM and with an additional dimension-eight operator
contribution following the EFT expansion of Eq.~(\ref{eq:eft}). In
particular, we set the coefficient  $f_{T,0}/\Lambda^4=0.1$/TeV${}^4$, while all other non-SM contributions are assumed
to vanish. New-physics contributions characterized by a coefficient of that size would not give rise to experimentally detectable signatures at the 13~TeV LHC. It is thus particularly interesting to see which  impact they have on observables at a 100~TeV FCC. 

As expected, deviations from the SM mostly occur at high scales. At
scales far above $M_{T_{WW}}\sim 2$~TeV, a naive implementation of
contributions from dimension-eight operators gives rise to an unphysical
rise of the distribution. This behavior can be avoided by a form
factor suppressing contributions beyond the reach of validity of the
EFT expansion. Requiring the VBS cross sections to preserve unitarity
at all scales, a suppression factor of the form specified in
Eq.~(\ref{eq:ff}) with the scale $\Lambda_F=4.7\TeV$ 
provides sufficient damping.  
The actual value of this scale factor has been obtained with the help of the calculator \cite{unitarity-diplom} available from \cite{unitarity-calc}. 

The impact of other dimension-eight operators on the transverse-mass distribution of the gauge-boson system is less pronounced. To enhance their relative impact on observables, we impose an additional transverse-mass cut of 
\beq
\label{eq:wpwp-cut}
M_{T_{WW}}>2~\mr{TeV}\,.
\eeq
Such a cut reduces the VBS cross section from $\sigma_\mr{SM}=49.34$~fb in the framework of the SM and $\sigma_{T,0}=49.52$~fb with the additional EFT contribution of the $\mc{O}_{T,0}$ operator with the coefficient and form factor suppression given above 
to  $\sigma_\mr{SM}^\mr{cut}=0.48$~fb and $\sigma_{T,0}^\mr{cut}=0.66$~fb, respectively. Obviously, the application of a severe cut on the gauge-boson system significantly improves the relative impact of the EFT contribution on the VBS cross section, although clearly the event rate at such high transverse-mass scales is small such that considerable integrated luminosity will be required to achieve a significant signature. 

\begin{figure}
\includegraphics[angle=0,width=0.5\textwidth]{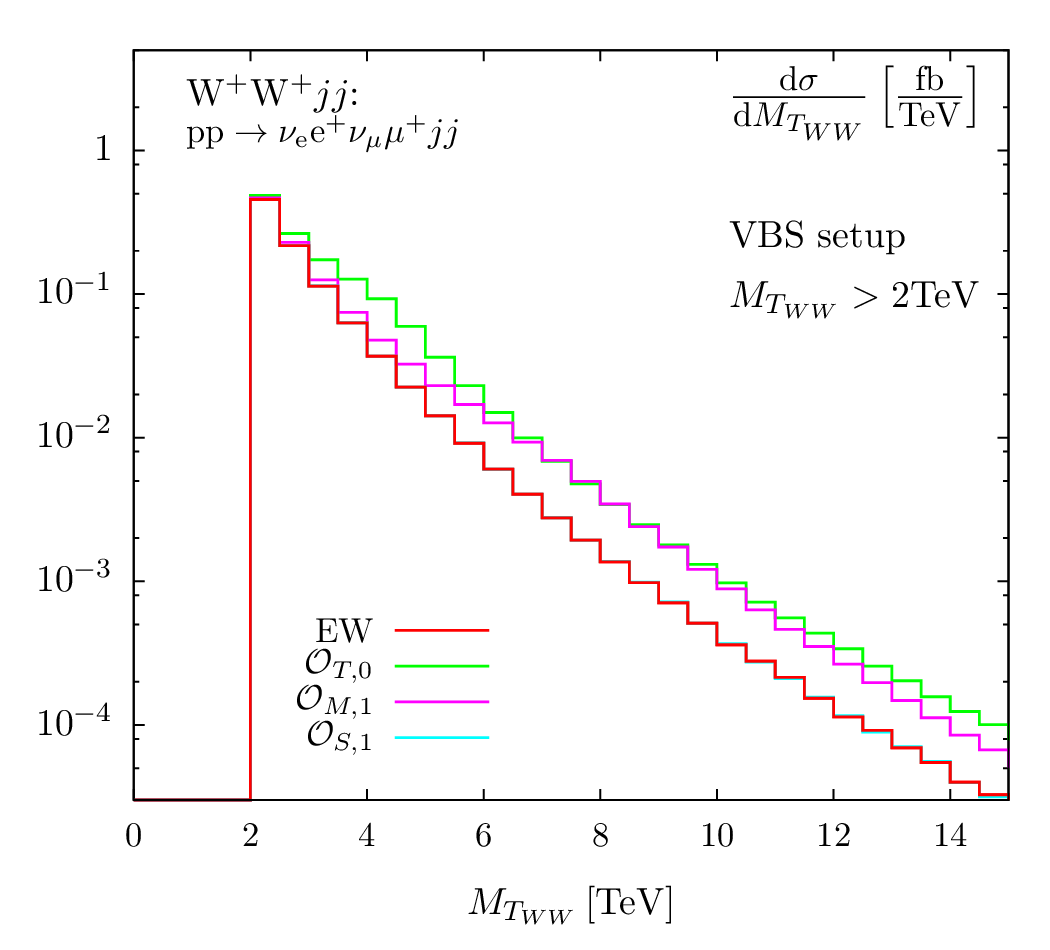}
\includegraphics[angle=0,width=0.5\textwidth]{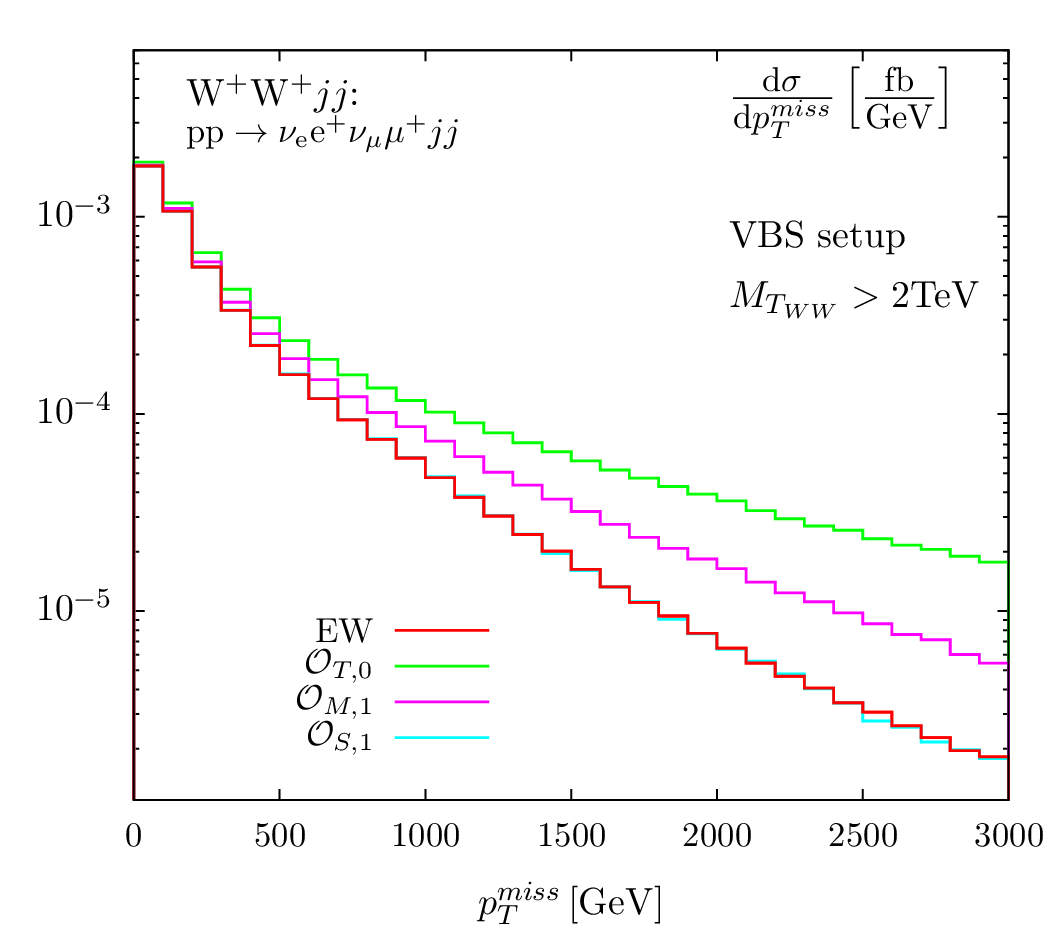}
\caption{\it 
Transverse-mass (l.h.s.) and missing transverse momentum distributions (r.h.s) for
the process $pp\to \nu_e e^+\nu_\mu \mu^+ jj$ in the framework of the
SM (red lines) and including the impact of the dimension-eight operators 
$\mc{O}_{S,1}$ (cyan lines), $\mc{O}_{M,1}$ (magenta lines), and $\mc{O}_{T,0}$  (green lines) supplied by an appropriate form factor. The selection cuts of
Eqs.~(\ref{eq:jet-def})--(\ref{eq:lgap}), Eq.~(\ref{eq:jjcuts_wpp}), and the additional transverse-mass cut of  Eq.~(\ref{eq:wpwp-cut})
are imposed. 
\label{fig:wpwpjj-agc-cut}
}
\end{figure}

In Fig.~\ref{fig:wpwpjj-agc-cut}
%
%
the transverse-mass and missing transverse-momentum distributions for our default $\wpwpjj$ settings with the additional cut of Eq.~(\ref{eq:wpwp-cut}) are shown separately for the SM and for simulations where either of the representative operator contributions $\mc{O}_{S,1}$, $\mc{O}_{M,1}$, or  $\mc{O}_{T,0}$ has been taken into account. In each case, the respective coefficient $f_{S,1}/\Lambda^4$, $f_{M,1}/\Lambda^4$, or $f_{T,0}/\Lambda^4$ has been set to $0.1$/TeV${}^4$, with all other EFT coefficients set to zero. To avoid unphysical unitarity violations at high energies, we apply form factors using the functional form of Eq.~(\ref{eq:ff}) with $\Lambda_F=4.3\TeV$ for the $\mc{O}_{S,1}$ operator, $\Lambda_F=8.3\TeV$ for $\mc{O}_{M,1}$, and -- as above -- $\Lambda_F=4.7\TeV$ for $\mc{O}_{T,0}$. The values of all scale factors have been fixed with the help of \cite{unitarity-diplom}. 

Clearly, a non-vanishing contribution from the $\mc{O}_{T,0}$ operator has the largest impact on each distribution. The operator $\mc{O}_{M,1}$ gives rise to slight enhancements in the tails of both considered distributions, while the impact of the $\mc{O}_{S,1}$ operator is barely distinguishable from the SM prediction. 
In the following, we will only consider contributions of the $\mc{O}_{T,0}$ operator. 

\begin{figure}
\includegraphics[angle=0,width=0.5\textwidth]{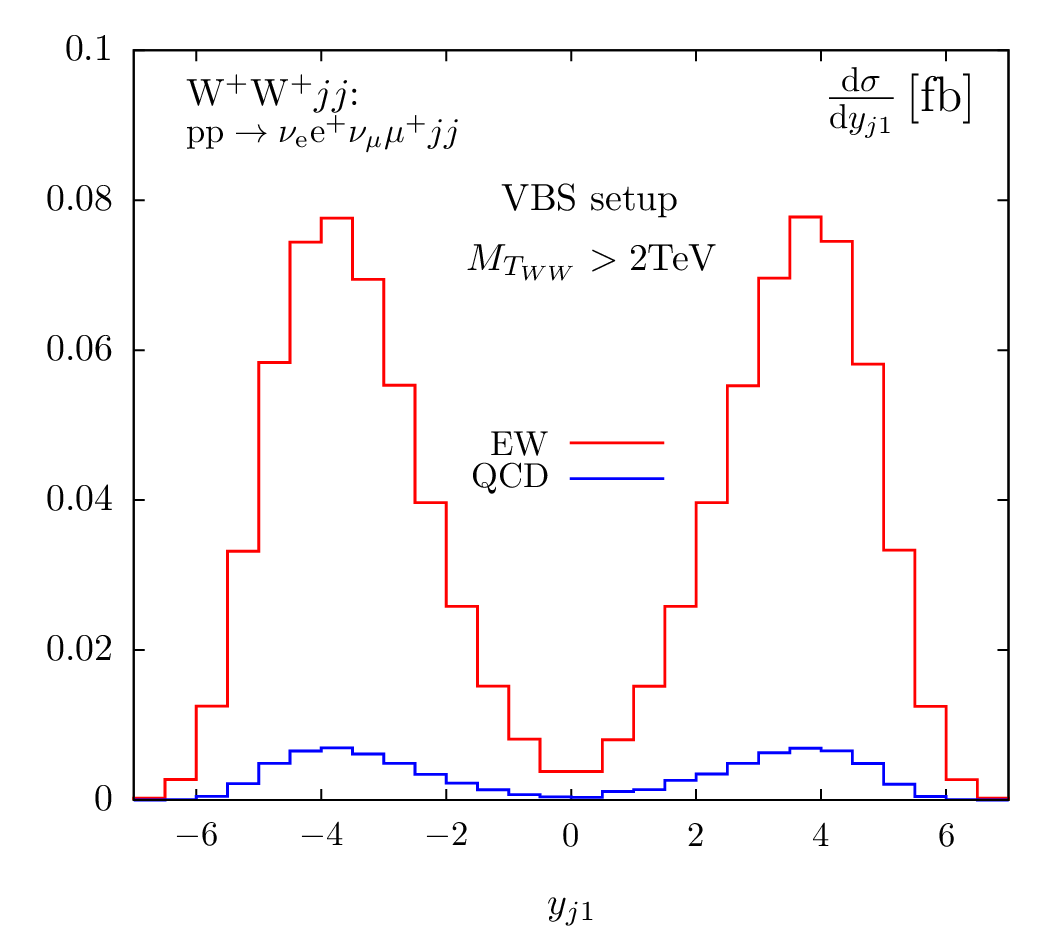}
\includegraphics[angle=0,width=0.5\textwidth]{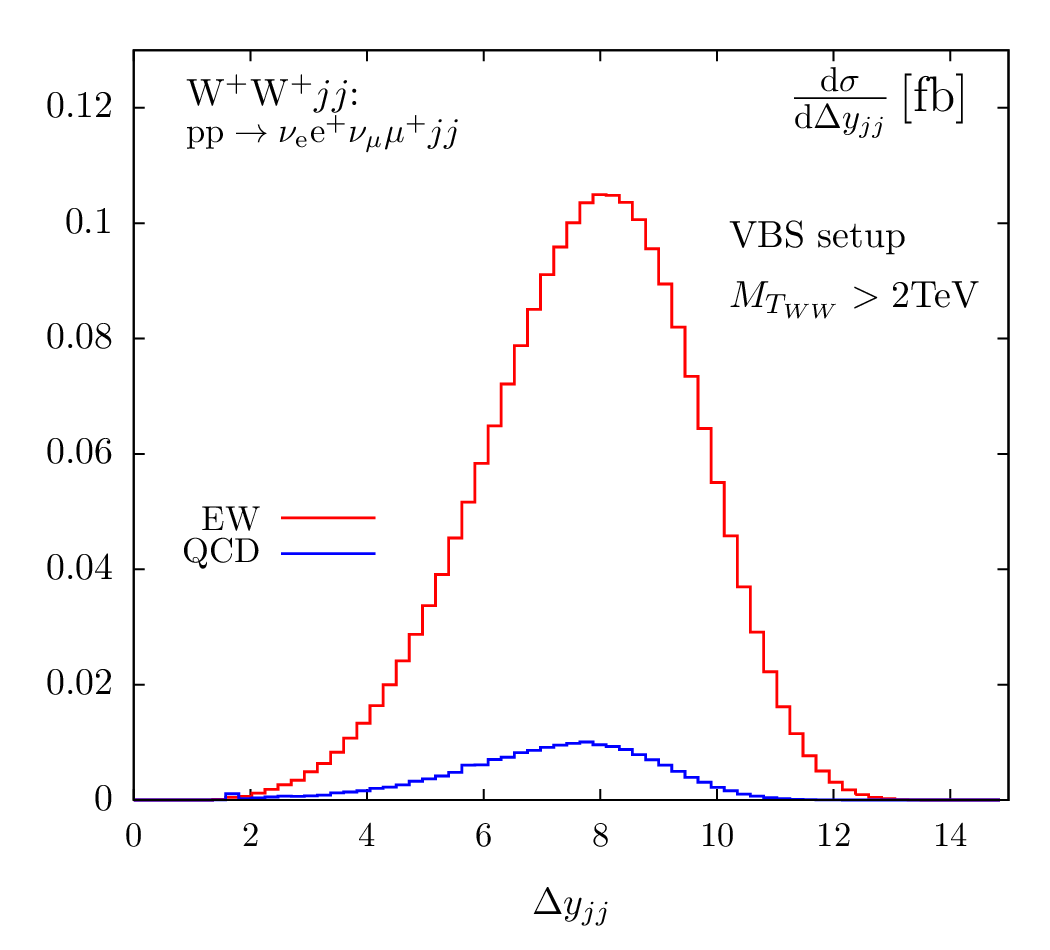}
\caption{\it Rapidity distribution of the hardest tagging jet~(l.h.s) and rapidity separation of the two
tagging jets~(r.h.s)  for the EW-induced (red lines) and
QCD-induced (blue lines) contributions to $pp\to \nu_e e^+\nu_\mu
\mu^+ jj$, within the selection cuts of
Eqs.~(\ref{eq:jet-def})--(\ref{eq:lgap}),~(\ref{eq:jjcuts_wpp}), and the additional transverse-mass cut of  Eq.~(\ref{eq:wpwp-cut}). 
\label{fig:wpwpjj-rap-mtcut}
}
\end{figure}

We note that the application of the transverse-mass cut of Eq.~(\ref{eq:wpwp-cut}) in addition to the $\wpwpjj$   specific selection cuts of Eqs.~(\ref{eq:jet-def})--(\ref{eq:lgap}) and Eq.~(\ref{eq:jjcuts_wpp}) changes the cross section contributions of the EW and QCD production modes in the framework of the SM from the values without such a cut to $\sigma_\mr{EW}^\mr{cut}=0.48$~fb  and $\sigma_\mr{QCD}^\mr{cut}=0.04$~fb. 
The large transverse mass goes hand-in-hand with an even larger invariant mass of the $W^+W^+$ system, which requires more energetic incoming quarks, on average. This induces a higher invariant mass of the tagging jet pair and, since the jet $p_T$'s remain of the order $m_W$, the tagging jets also are found at somewhat larger rapidity. This is illustrated 
for the rapidity distribution of the hardest jet and the rapidity separation of the two tagging jets in Fig.~\ref{fig:wpwpjj-rap-mtcut}, 
%
%
which are to be compared to the respective distributions without the $M_{T_{WW}}$ cut, shown 
in Figs.~\ref{fig:wpwpjj-jets} and \ref{fig:wpwpjj-cuts}. The slightly higher tagging jet rapidities are also reflected in a larger rapidity separation of the two tagging jets which now peaks  around $\Delta y_{jj}\approx 8$  for the EW mode  
(as compared to about 6.5 before the transverse-mass cut). A similar trend of larger rapidities occurs for the QCD background. Since the event rates  associated with the QCD production mode are very small, however, this does not pose a serious issue for the separation of signal and background in this high-invariant mass regime.  
Note that excellent forward rapidity coverage of the hadron calorimeters is needed at the FCC in order not to loose a substantial fraction of VBS events in this particularly interesting region of high invariant mass of the vector boson pair. 

%
\subsection{$\wpzjj$}
\label{sec:wpzjj}

In contrast to the $\wpwpjj$ channel that is free of gluon-induced
background processes, all other VBS reactions are plagued by large QCD
backgrounds that require more powerful cuts on the tagging-jets system
than those that have been considered in the $\wpwpjj$ mode.

To quantify this statement, let us consider the representative $pp\to
e^+ \nu_e \mu^+ \mu^- jj$ final state for the $\wpzjj$ channel. With
the basic cuts of Eqs.~(\ref{eq:jet-def})--(\ref{eq:ll-cuts}), the
associated QCD background cross section, 
$\sigma_\mr{basic}^\mr{QCD} =23.38$~fb,
by far overshoots the EW signal cross section,
$\sigma_\mr{basic}^\mr{EW} =8.46$~fb,
resulting in a 
$S/B$ ratio of less than 0.4. 
Besides the basic cuts of
Eqs.~(\ref{eq:jet-def})--(\ref{eq:ll-cuts}) severe additional cuts on
the tagging jets' invariant mass and rapidity separation need to be
applied to significantly reduce the impact of the QCD
background.  In particular, we require 
\begin{equation}
\label{eq:jjcuts_wpz}
M_{jj}>2500~\mathrm{GeV}\,,\quad \quad \quad \quad
\Delta y_{jj}>5\,.
\end{equation}
 Indeed, with these customized selection cuts, the VBS cross section
is reduced by merely about $40\%$ with respect to the value obtained
with basic selection cuts only, while the QCD cross section goes down
by a factor of 8.4.  
As a consequence,  
one arrives at  signal and background cross sections of, respectively, 
 $\sigma_S= 5.08$~fb and
$\sigma_B= 2.79$~fb, resulting in a $S/B$ ratio of about 1.8. 

\begin{figure}
\includegraphics[angle=0,width=0.5\textwidth]{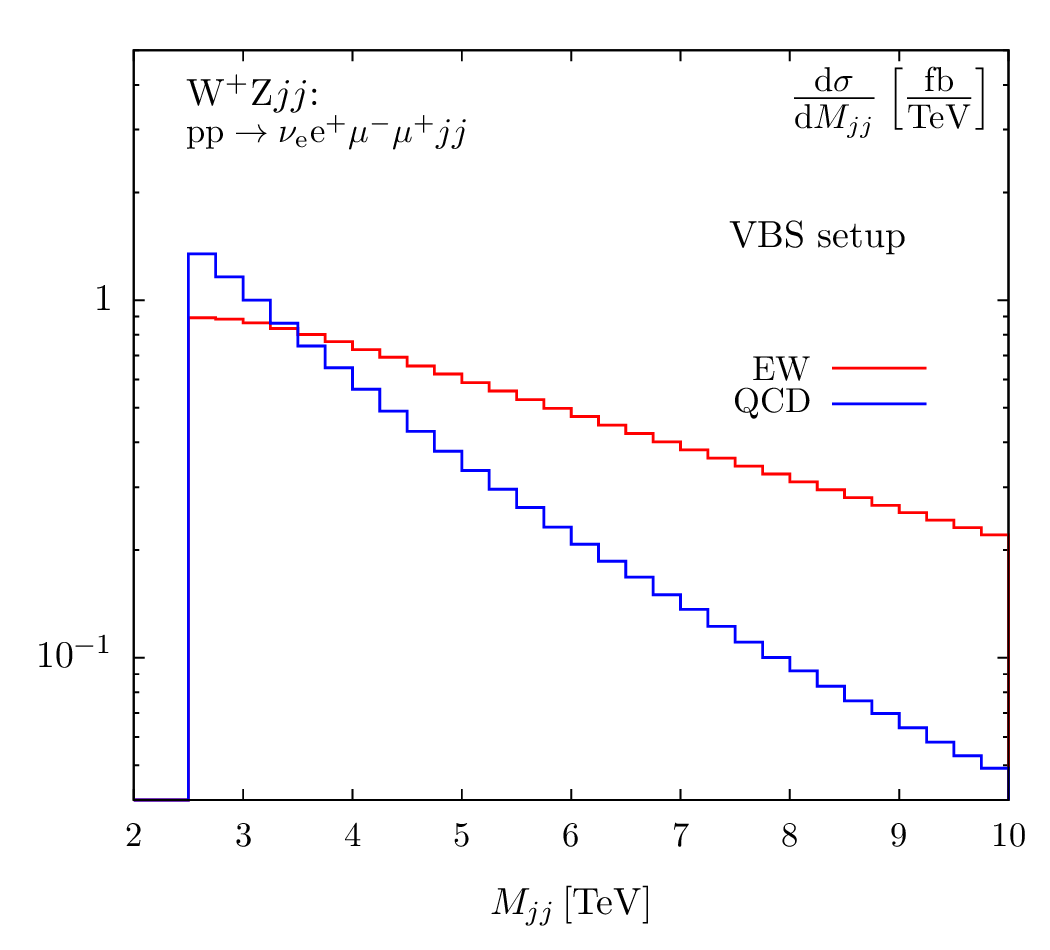}
\includegraphics[angle=0,width=0.5\textwidth]{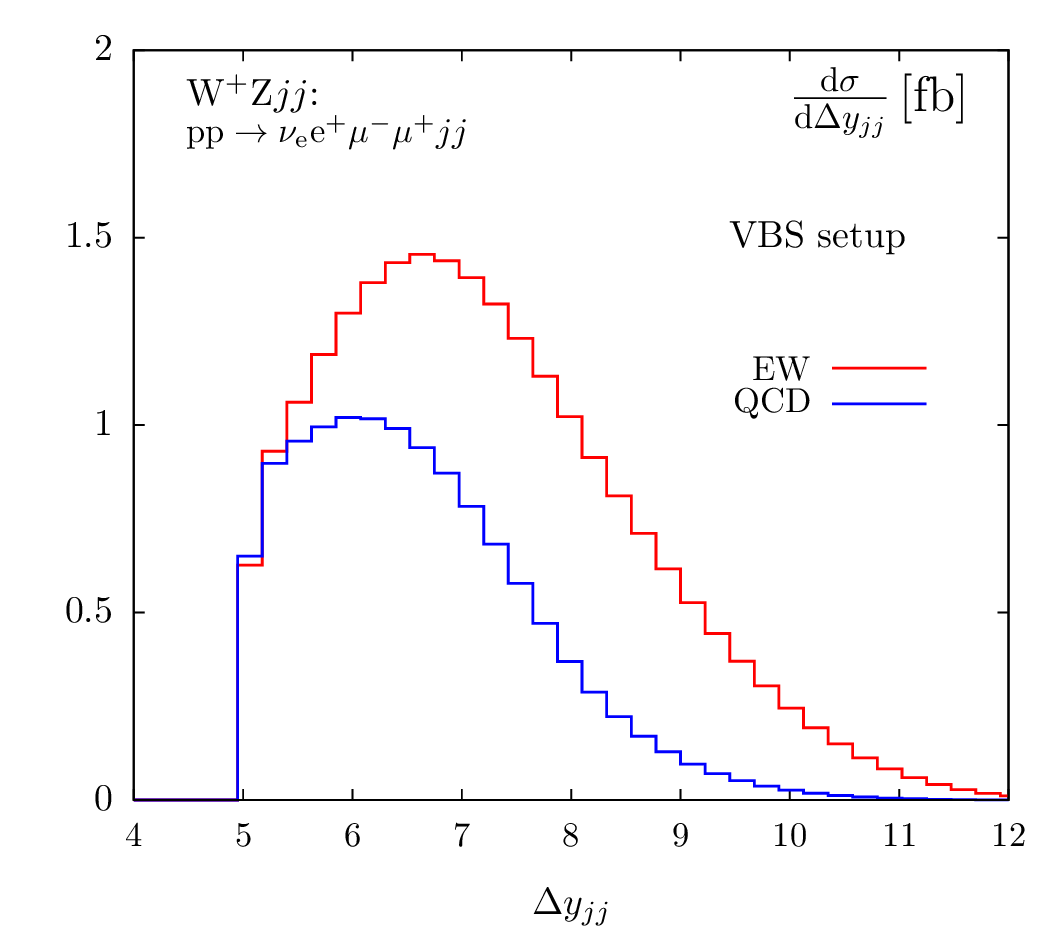}
\caption{\it Invariant mass~(l.h.s.) and rapidity separation of the two
tagging jets~(r.h.s) for the EW-induced (red lines) and QCD-induced
(blue lines) contributions to $pp\to \emmvjj$, within the
selection cuts of Eqs.~(\ref{eq:jet-def})--(\ref{eq:ll-cuts}) and
Eq.~(\ref{eq:jjcuts_wpz}).
\label{fig:wzjj-cuts}
}
\end{figure}

\begin{figure}
\center
\includegraphics[angle=0,width=0.5\textwidth]{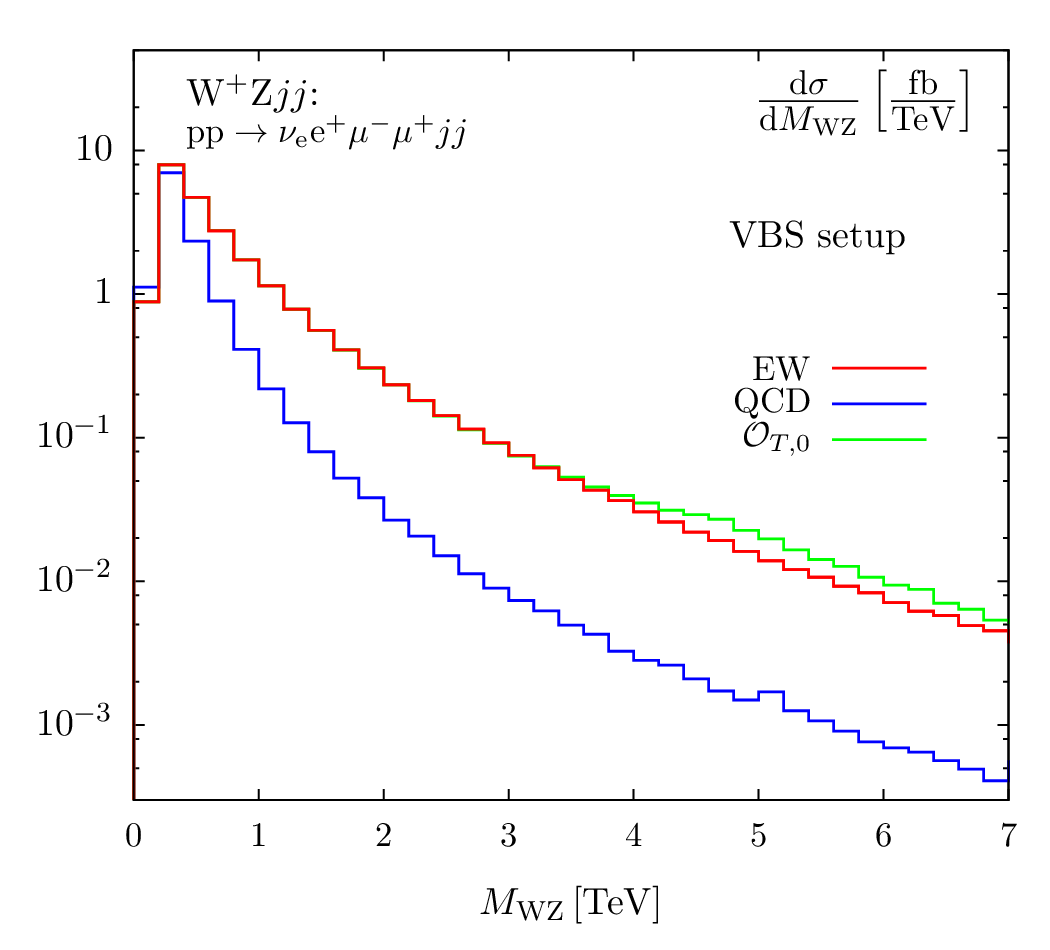}
\caption{\it Invariant-mass distribution of the $WZ$ system reconstructed
from the lepton momenta for the process  $pp\to \emmvjj$. Shown are the 
QCD-induced contributions (blue line), and the EW-induced contributions in the framework of the SM (red line) and including the impact of the dimension-eight operator 
$\mc{O}_{T,0}$ (green line) with the form
factor of Eq.~(\ref{eq:ff}). 
The selection cuts of
Eqs.~(\ref{eq:jet-def})--(\ref{eq:ll-cuts}) and
Eq.~(\ref{eq:jjcuts_wpz}) are applied. 
\label{fig:wpzjj-mvv}
}
\end{figure}

Figure~\ref{fig:wzjj-cuts} illustrates clearly the different behavior
of the dijet system in the $\wpzjj$ mode compared to the $\wpwpjj$
channel considered in Figure~\ref{fig:wpwpjj-cuts}.
%
%
While in the $\wpwpjj$ channel the VBS contribution exceeds the QCD
contribution in the entire plotted range of  $M_{jj}$, in
the $\wpzjj$ mode this is the case only in the tail of the
distribution.   Already at around $M_{jj}\sim 3.5$~TeV
the QCD contribution starts to dominate and steeply rises towards
smaller values of the invariant mass.  Even though the $pp \to
e^+\nu_e \mu^+ \mu^- jj$ final state contains a neutrino that is
invisible to the detector, the invariant-mass distribution of the
 leptonic decay products, $M_{WZ}$, can be reconstructed, if kinematic
constraints are used to determine the longitudinal component of the
neutrino momentum.  In the
framework of the SM, there exists no particle with appropriate quantum
numbers and mass to resonantly produce an on-shell $W^+Z$ system.  The
invariant-mass distribution thus does not exhibit pronounced resonance
peaks, but is characterized by a broad continuum, as can be
  observed in Fig.~\ref{fig:wpzjj-mvv} 
%
%
 in the framework of the SM and with an additional dimension-eight operator
contribution following the EFT expansion of Eq.~(\ref{eq:eft}). To simulate the latter, we set the coefficient  $f_{T,0}/\Lambda^4=0.1$/TeV${}^4$ and supply a suppression factor of the form specified in Eq.~(\ref{eq:ff}) with the scale $\Lambda_F=4.7\TeV$. All other non-SM contributions are assumed to vanish.
Above $M_{WZ}=4$~TeV one finds 
$\sigma_\mr{SM}^\mr{cut}=0.047$~fb for the SM and $\sigma_{T,0}^\mr{cut}=0.061$~fb in the presence of the $T_0$ coupling. With sufficient integrated luminosity the difference will be highly significant.

Distinct peaks in the $M_{WZ}$ spectrum are expected in BSM models  
featuring new particles that can resonantly decay into the 
$WZ$ system. The most prominent examples comprise models with a
$W^\prime$ boson or Kaluza-Klein models where the compactification of
an extra space-time dimension gives rise to a tower of new gauge
bosons with masses larger than a few hundred GeV. The impact of such
new resonances on VBS signatures,  in the context of the LHC, was
discussed for example  in \cite{Englert:2008tn}.

%
\subsection{$\zzjj$}
\label{sec:zzjj}
%

Compared to VBS processes involving $W$~bosons, relatively small event
rates are encountered for $\zzjj$ VBS production, in particular in the
fully leptonic decay modes. This is partly due to the fact that the
electroweak couplings of $Z$~bosons to quarks that occur in the
associated VBS production process are smaller than respective
couplings involving $W$~bosons. Moreover, the branching ratios of the
$Z$~bosons to charged leptons are approximately three times smaller
than the branching ratios of the $W$~bosons into $\ell\nu$ pairs.
Nonetheless, $\zzjj$ reactions provide a very clean testbed for VBS
processes since the kinematics of the $ZZ$ system is fully
reconstructible, if final states with four charged leptons are
considered. We will thus focus on the $pp\to \eemmjj$ process in the
following. We note that, although we simplistically refer to the
process under consideration as $\zzjj$ production, QCD and EW
production of $\eemmjj$ final states include also Feynman
diagrams with  the exchange of photons instead of $Z$-bosons,  and respective interference
contributions. These diagrams fully have to be taken into account for
the computation of meaningful cross sections, in order not to violate
electroweak gauge invariance. Potentially divergent contributions
involving a $\gamma^* \to\ell^+\ell^-$ splitting at very low photon
virtuality are removed, however, by the invariant-mass cuts of
Eq.~(\ref{eq:ll-cuts}). In order to suppress contributions from
QCD-induced $\eemmjj$ production, we require
\begin{equation}
\label{eq:jjcuts_zz}
M_{jj}>2000~\mathrm{GeV}\,,\quad \quad \quad \quad
\Delta y_{jj}>3\,, 
\end{equation}
in addition to the basic cuts of
Eqs.~(\ref{eq:jet-def})--(\ref{eq:ll-cuts}).  With these cuts we
have obtained the cross sections
$\sigma_S = 2.18$~fb and $\sigma_B = 0.23$~fb
for the signal and
background processes, respectively,  resulting in  the S/B ratio of 
almost $10$. 

\begin{figure}
\includegraphics[angle=0,width=0.5\textwidth]{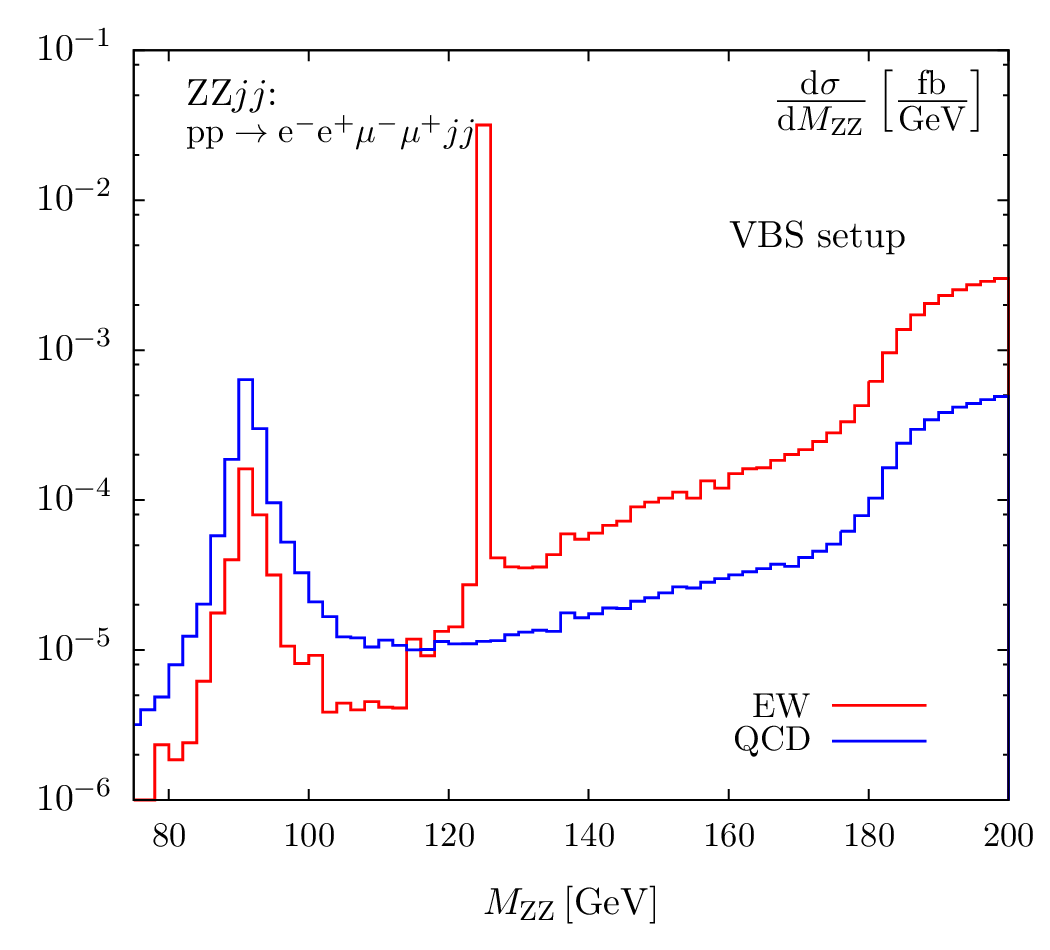}
\includegraphics[angle=0,width=0.5\textwidth]{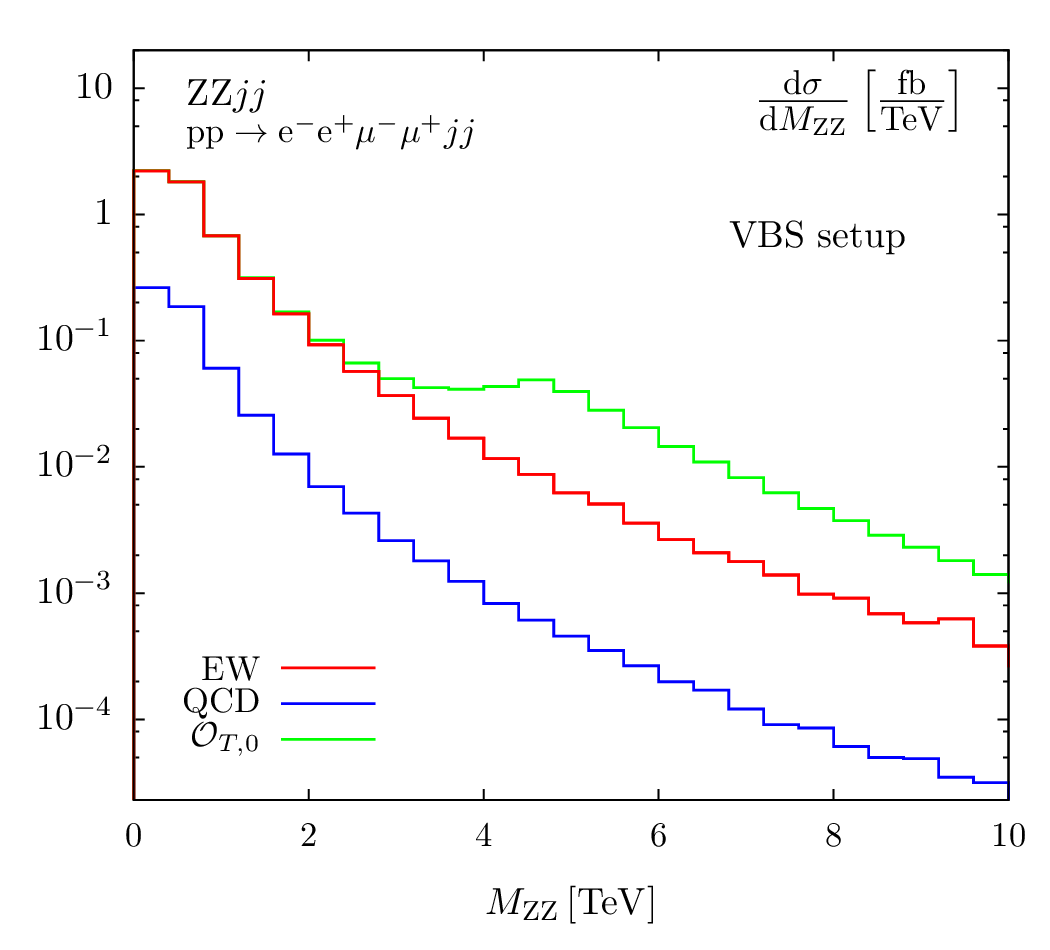}
\caption{\it 
Invariant-mass distribution of the $ZZ$ system reconstructed
from the lepton momenta in two different plot ranges for the
process $pp\to \eemmjj$ in the framework of the SM (red and blue lines) and
including the impact of the dimension-eight operator $\mc{O}_{T,0}$ 
(green line) with the form factor of
Eq.~(\ref{eq:ff}). The selection cuts of
Eqs.~(\ref{eq:jet-def})--(\ref{eq:ll-cuts}) and
Eq.~(\ref{eq:jjcuts_zz}) are imposed. %
\label{fig:zzjj-mvv}
}
\end{figure}

For the EW signal process the
invariant-mass distribution of the four-lepton system,
$d\sigma/dM_{ZZ}$, exhibits a rich resonance structure, featuring
peaks at $m_{4\ell}=M_{ZZ}=m_Z$, and at the Higgs resonance $M_{ZZ}\approx m_H$, 
as can be seen in
Fig.~\ref{fig:zzjj-mvv}.
%
%
Similar to the $\wpwpjj$ channel discussed before, we explore the
impact of the dimension-eight operator $\mc{O}_{T,0}$ on the
$\zzjj$ reaction, setting the EFT coefficient
$f_{T,0}/\Lambda^4=0.1$/TeV${}^4$, and assuming all other non-SM
contributions to be negligible.  To warrant unitarity conservation up
to the highest scales, we use the form factor of Eq.~(\ref{eq:ff})
with $\Lambda_{F}=4.7\TeV$.  For the VBS cross section in the context
of the SM and the respective EFT result within the $ZZjj$-specific
selection cuts of Eqs.~(\ref{eq:jet-def})--(\ref{eq:ll-cuts}) and
Eq.~(\ref{eq:jjcuts_zz}), we find results that are rather similar
to each other, $\sigma_{SM} = 2.18$~fb and $\sigma_\mr{EFT}= 2.30$~fb. 
As becomes obvious from an inspection of the invariant mass
distribution in the TeV range, shown in the right panel of
Fig.~\ref{fig:zzjj-mvv}, differences between the SM and the EFT
prediction are most pronounced at high scales.  Indeed, by applying an
additional selection cut of
\beq
M_{ZZ} > 2~\mr{TeV}\,,
\eeq
the relative difference between the VBS cross section in the context
of the SM and the respective EFT result can be significantly enhanced,
yielding $\sigma_\mr{SM}^\mr{cut} = 0.11$~fb and
$\sigma_\mr{EFT}^\mr{cut} = 0.22$~fb,   
i.e.\ even anomalous couplings substantially smaller than 0.1~TeV$^{-4}$ can be seen in this channel with 3000~fb$^{-1}$. 
Note that in the high-invariant mass region most sensitive to the EFT operator the impact of the QCD-induced background is negligible.

%
\subsection{$\wpwmjj$}
\label{sec:wpwmjj}
%

The $\wpwmjj$ mode exhibits the largest cross section of all VBS
channels. Nonetheless, extracting the VBS signal from the background
is particularly challenging in this mode, because in addition to the
QCD-induced $\wpwmjj$ contribution very large background rates arise
from  top-quark pair production processes.  First, in reactions
of the type $pp\to t\bar t$,  almost all top quarks decay into
$W$~bosons and bottom quarks. The bottom quarks may be misidentified
experimentally as light-flavor tagging jets. Together with the decay
products of the $W$~bosons, such bottom-jets may give rise to
signatures very similar to the VBS signal process.  Second, the light
jets occurring in $t\bar t +\mr{jets}$ production processes may mimic
the tag jets characteristic of VBS reactions. Because of the large
production rates associated with these  top-quark induced
production processes, they remain a significant source of background even
after the application of stringent selection cuts.

For our analysis, we focus on the $\evmvjj$ final state.  In
addition to the QCD-induced $pp\to \evmvjj$ background, we consider
the top-quark pair production process, $pp\to \evmvbb$, together with the 
associated production of a top-quark pair with one and two jets,
i.e.\  $pp\to \evmvbbj$ and $pp\to \evmvbbjj$, where $j$ stands for a
light quark or a gluon.  All top-quark induced background processes are simulated
with the \HELAC{} package where all double-, single and non-resonant
contributions, interferences and off-shell effects due to the finite
width of the top quarks and $W$~bosons are fully incorporated in the
matrix element calculations.  In case of the top-quark pair production
process both massive bottom quarks from top-quark decays are
identified as the tagging jets. We will denote this process as $t\bar
t$ production. In case of the $\evmvbbj$ and $\evmvbbjj$ processes
various phase-space regions can be studied. First, for $\evmvbbj$ we
can identify the phase-space region where both bottom quarks are again
associated with the two (hardest) tagging jets whereas an additional
light quark or gluon is not restricted by any cuts.  This configuration is,
however, properly described only by next-to-leading-order
calculations like the one described, for example, in
Ref.~\cite{Bevilacqua:2010qb}. Since we are interested in LO results 
only we do not consider higher order corrections to the $t\bar{t}$
production process and discard this configuration.  Nevertheless,
there is still another distinctly different region of phase space,
which can be studied at LO, namely the one where the final state light
quark or gluon can give rise to one tagging jet, and one of the two
bottom jets is identified as the other tagging jet. We shall call this
configuration $t\bar{t}j$ production. In the next step, we consider
$\evmvbbjj$ production. Once more we can identify two massive bottom
quarks as the two tagging jets and leave additional partons
unrestricted. This is, however, a next-to-next-to-leading-order
correction to $t\bar{t}$ production. We can also identify only one bottom
quark with one tagging jet and the second tagging jet with a light quark
or gluon. This is, however,  a part of the next-to-leading-order corrections to
our $t\bar{t}j$ production process, where results from
Ref.~\cite{Bevilacqua:2015qha,Bevilacqua:2016jfk} could be of
help. We disregard these cases and keep only the new configuration
where the final state light  quarks  or gluons are the two tagging
jets and  both bottom quarks are unconstrained. We denote this part of
the top-quark background as $t\bar{t}jj$. In each case two tagging jets
are subject to the cuts of Eqs.~(\ref{eq:jet-def})--(\ref{eq:lgap}). 

Discrimination between jets originating from $b$-quarks and those
emerging from light quarks or gluons can be very helpful to further
suppress  $t\bar{t}+\mr{jets}$ backgrounds in the $\wpwmjj$
channel. The top-induced background contributions can be suppressed
most efficiently by removing all events that contain an identified
bottom jet. To this end a $b$-tagging jet veto is applied to
eliminate any events where at least one of the tagging jets is
identified as arising from a $b$-quark or anti-quark. The procedure can be
applied to both the $t\bar{t}$ and $t\bar{t}j$ production processes. However, the
$b$-tagging efficiency is not ideal. Since we are not aware of a respective 
study for future $100$~TeV $pp$ colliders we concentrate on related 
results for the LHC to estimate the impact of the $b$-jet tagging efficiency on the
$b$-tagging jet veto. To this end we incorporate the CMS analysis of Ref.~\cite{Weiser:2006md} for our assumptions on $b$-tagging efficiencies
and mis-tagging probabilities.  To be more precise, we assume that the
$b$-tagging efficiency depends on the transverse momentum and rapidity
of the respective tagging jet as detailed in Tab.~\ref{tab:btag}. 
%
\begin{table}
\begin{center}
\begin{tabular}{|c|c|c|}
\hline
& &\\
$p_\mr{T,jet}$ [GeV] & $1.4 < |y_j|$  & $|y_j|<1.4$
  \\
& &\\
\hline
& & \\
50 - 80   & $65\%$ & $75\%$ \\
80 - 120  & $70\%$ & $80\%$ \\
120 - 170 & $70\%$ & $80\%$ \\
$>$ 170   & $65\%$ & $75\%$ \\
& &\\
\hline
\end{tabular}
\caption{\it
\label{tab:btag}
Assumed $b$-tagging efficiencies as functions
of the transverse momentum of the jet for different rapidity ranges
 (adapted from \cite{Englert:2008tn}).}
\end{center}
\end{table}
%
To diminish the top-quark background even further one can 
make use of the so-called central-jet veto.  For the
$t\bar{t}j$ and $t\bar{t}jj$ backgrounds one or
both of the $b$-quarks are not identified as the tagging jets. 
  They
will most frequently lie between the two tagging jets in rapidity,  i.e.\  in the region
where we look for the decay products of two gauge bosons. Vetoing
events with  these additional $b$ jets provides a powerful suppression
tool to control the top-quark background.  Thus, we reject all events where a
$b$~jet with $p_{T,b}  > 50$ GeV is observed in the rapidity gap region between
the two tagging jets,  
\begin{equation}
\label{eq:jveto}
 y_{j,min}^\mr{tag} < y_\mr{jet}  < y_{j,max}^\mr{tag}\,. 
\end{equation}
We note that extra jet activity does not require a bottom jet, but merely
the rejection of any events with an  additional jet.  In our leading-order
simulation, with leptonic decays of both gauge bosons, this
central-jet veto criterion only affects the $t\bar{t}j$ and
$t\bar{t}jj$ backgrounds. At this perturbative order, the $t\bar{t}$
production process, the QCD-induced and the VBS-type $\wpwmjj$
contributions exhibit at most two hard jets that serve as tagging
jets.  However, additional jet activity would occur also in these
reactions in higher-order calculations that include further
parton-emission contributions. Moreover, hadronic decays of one or
both gauge bosons would contribute to the multiple production of jets in
these processes.  It is a well-known feature of VBS processes,
however, that in contrast to the case of QCD-induced reactions extra
parton emission typically occurs close to or more forward than the tagging jets and thus
gives rise to little jet activity in the central-rapidity
region~\cite{Rainwater:1999sd,Kauer:2000hi,Figy:2007kv,
Campanario:2013fsa,Jager:2014vna}.  In addition to the basic cuts of
Eqs.~(\ref{eq:jet-def})--(\ref{eq:lgap}), the $b$-tagging jet veto as
well as the central-jet-veto of Eq.~(\ref{eq:jveto}), the
process-specific cuts of 
\begin{equation}
\label{eq:jjcuts_wpwm}
M_{jj}>2000~\mathrm{GeV}\,,\quad \quad \quad \quad
\Delta y_{jj}>5\,, 
\end{equation}
are used. The cross sections obtained in this way can be summarized as
follows: $\sigma_S=58.27$~fb, $\sigma_B^\mr{QCD}=22.26$~fb and
$\sigma_B^{t\bar{t}+jets}=589$~fb ($\sigma_B^{t\bar{t}}=0.1528$ fb,
$\sigma_B^{t\bar{t}j}=84.24$ fb and $\sigma_B^{t\bar{t}jj}=505$
fb). These numbers yield a $S/B$ ratio of $0.09$, which obviously would make an extraction of the signal very difficult. 
To improve the
signal to background ratio we have thus applied an additional cut on the
transverse mass of the final state system reconstructed from the
momenta of the charged leptons and the missing transverse momentum due
to the neutrinos in the final state according to the definition of
Eq.~(\ref{eq:def-mtww}), 
\begin{equation}
\label{eq:wpwm-mt}
M_{T_{WW}} > 1 ~{\rm TeV}\,.
\end{equation}
With this cut the following cross section results have been obtained
instead: $\sigma_S=3.59$~fb, $\sigma_B^\mr{QCD}=0.389$~fb and
$\sigma_B^{t\bar{t}+jets}=4.23$~fb ($\sigma_B^{t\bar{t}}=0.02$~fb,
$\sigma_B^{t\bar{t}j}=0.76$~fb and $\sigma_B^{t\bar{t}jj}=3.45$~fb). This time the more advantageous $S/B$ ratio of $0.8$ has been reached.  

In order to investigate the impact  of the $p_{T,b}$ cut in the central-jet veto
procedure on the $\sigma_B^{t\bar{t}j}$ and
$\sigma_B^{t\bar{t}jj}$ cross sections, we have  relaxed its value  to  $p_{T,b} > 100$~GeV 
once the $M_{T_{WW}}$ cut has already been applied. With $p_{T,b} > 100$~GeV we
have obtained the following results: $\sigma_B^{t\bar{t}j}=1.05$~fb and
$\sigma_B^{t\bar{t}jj}=7.61$~fb, that together with
$\sigma_B^{t\bar{t}}=0.02$~fb result in  the following total cross
section for the top quark background process: 
 $\sigma_B^{t\bar{t}+jets}=8.68$~fb.  A difference by a factor of two with respect to the corresponding result in the setup  with a more severe transverse-momentum cut can
be observed -- the consequence of which is the now less advantageous $S/B$ ratio of
$0.4$. Thus, our final setup for the $\wpwmjj$ channel
comprises  the original  (harder) central jet-veto cut
of  $p_{T,b} > 50$~GeV as well as the set of selection cuts from 
 Eqs.~(\ref{eq:jet-def})--(\ref{eq:lgap}), 
Eqs.~(\ref{eq:jveto})--(\ref{eq:wpwm-mt}).

\begin{figure}
\includegraphics[angle=0,width=0.5\textwidth]{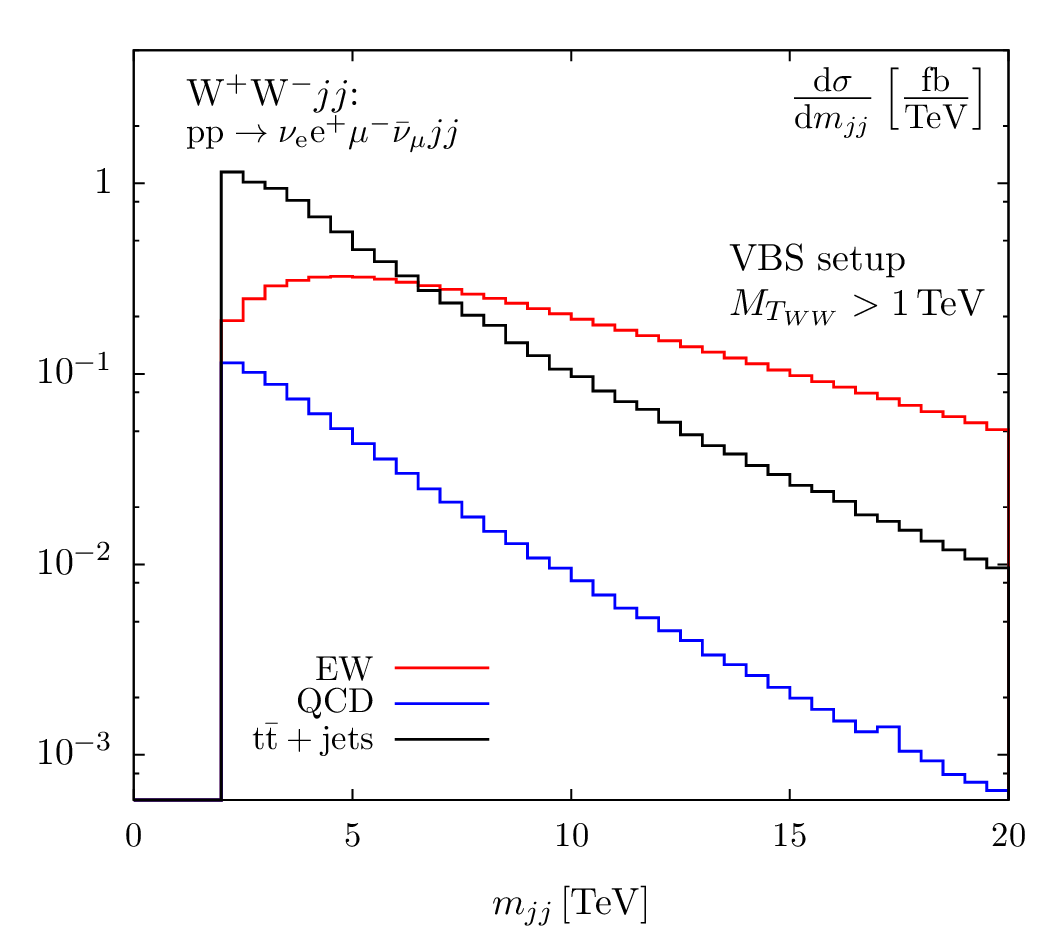}
\includegraphics[angle=0,width=0.5\textwidth]{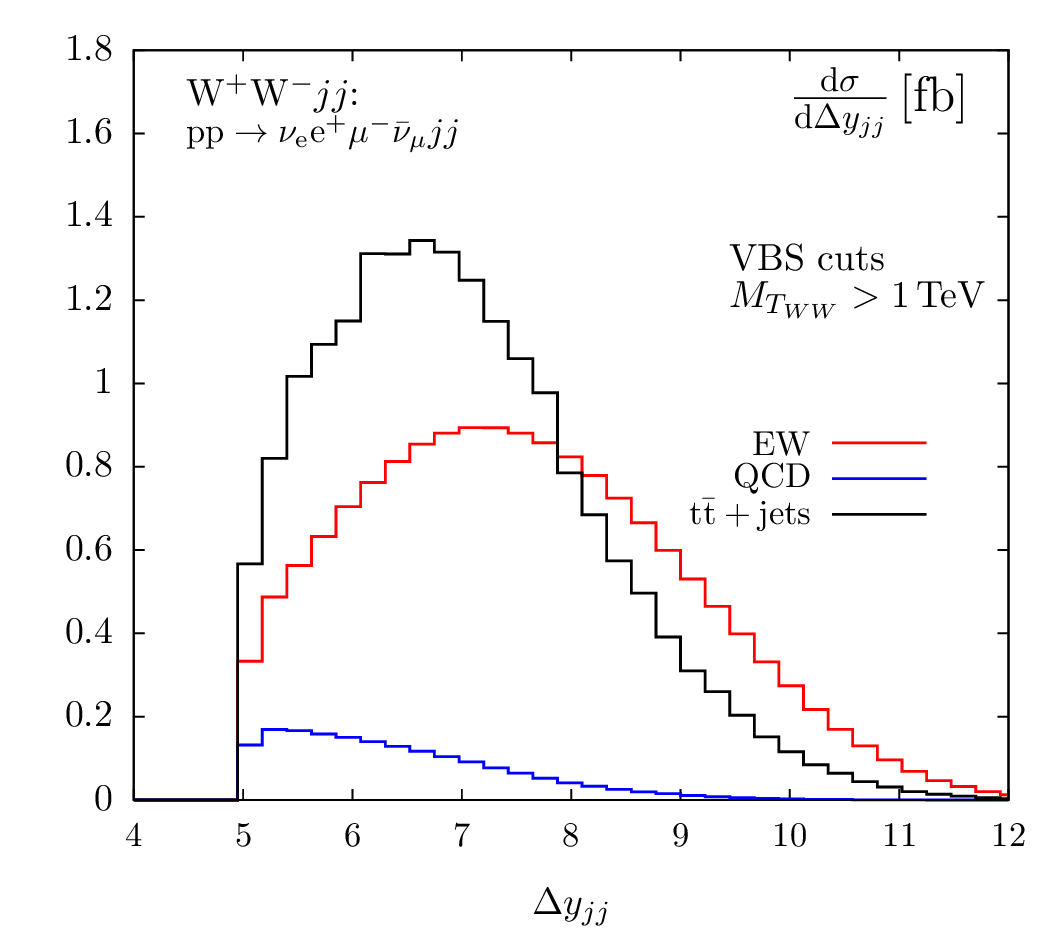}
\caption{\it Invariant mass (l.h.s.) and rapidity separation (r.h.s.)
  of the tagging jets 
for  the EW-induced signal (red lines), the QCD-induced (blue lines),
and the top-quark induced (black lines) background contributions to  $pp\to
 \evmvjj$, 
within the selection cuts of
Eqs.~(\ref{eq:jet-def})--(\ref{eq:lgap}) and
Eqs.~(\ref{eq:jveto})--(\ref{eq:wpwm-mt}).
\label{fig:wpwmjj-mjj}
}
\end{figure}

Within this setup, Fig.~\ref{fig:wpwmjj-mjj} illustrates the invariant-mass and rapidity distributions of the tagging jets. 
%
%
The signal cross section  dominates at high invariant mass, decreasing more slowly than the various background contributions. In the rapidity-separation variable, the VBS signal tends to larger values than the top backgrounds and, in particular, the QCD-induced $\wpwmjj$ contributions.  
Together with the transverse-mass cut of Eq.~(\ref{eq:wpwm-mt}) the cuts of Eq.~(\ref{eq:jjcuts_wpwm}) have removed large parts of the QCD- and top-quark-induced background contributions leaving us with an advantageous $S/B$ ratio due to the dominance of the VBS signal in the remaining regions of phase space.

\begin{figure}
\includegraphics[angle=0,width=0.5\textwidth]{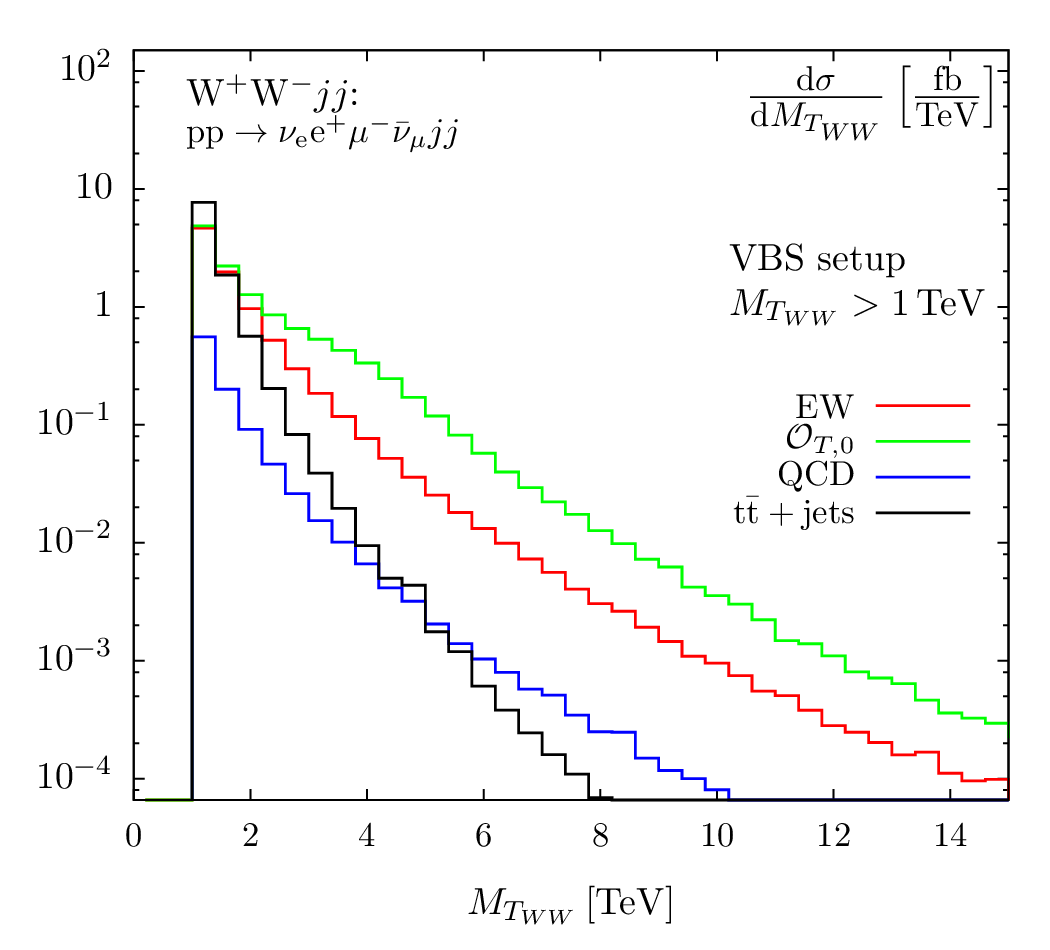}
\includegraphics[angle=0,width=0.5\textwidth]{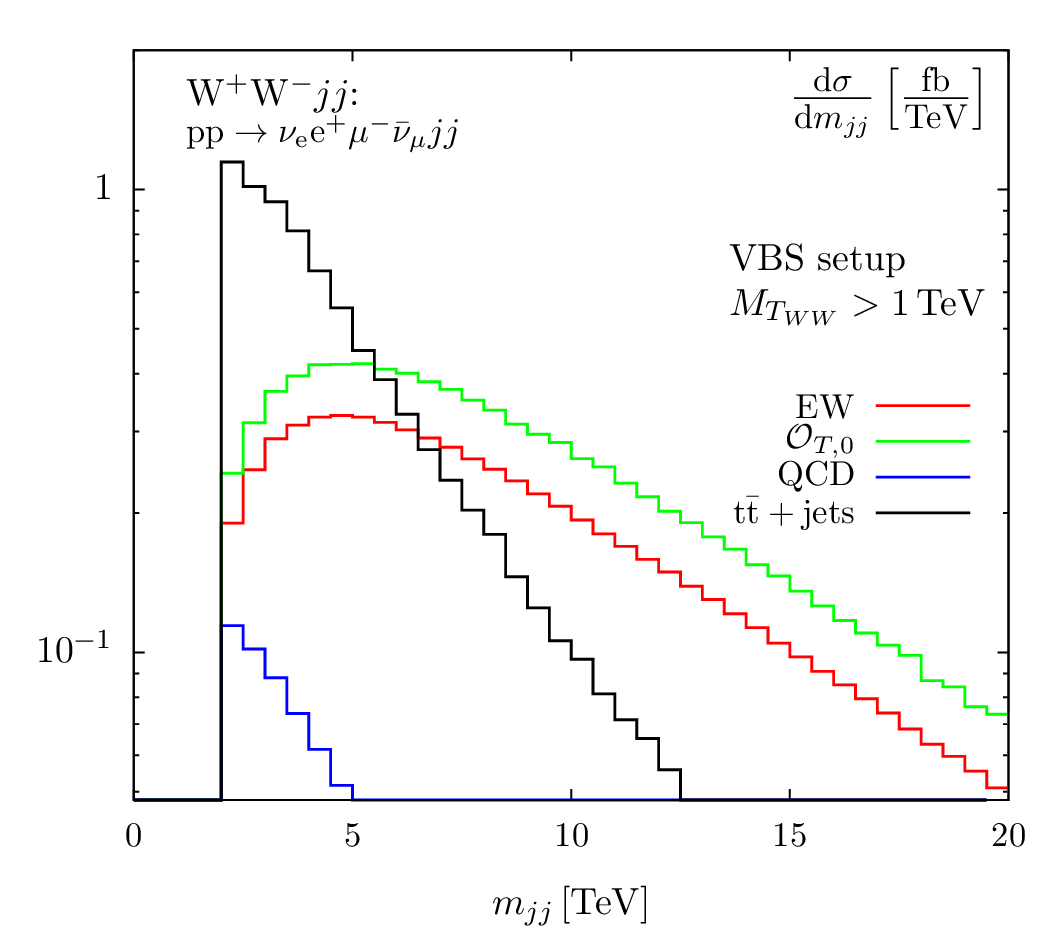}
\caption{\it  
Transverse mass of the $W^+W^-$ system (l.h.s) and invariant mass of the tagging jets (r.h.s) 
in the framework of the SM (red lines) and including the impact of the dimension-eight operator 
$\mc{O}_{T,0}$ (green lines) with  the form
factor of Eq.~(\ref{eq:ff}). Also shown are the the QCD-induced (blue lines),
and the top-quark induced (black lines) background contributions. 
The selection cuts of
Eqs.~(\ref{eq:jet-def})--(\ref{eq:lgap}) and Eqs.~(\ref{eq:jveto})--(\ref{eq:wpwm-mt}) are imposed in each case. 
\label{fig:wpwmjj-mt-eft}
}
\end{figure}

In analogy to the $\wpwpjj$ and $\zzjj$ channels discussed previously, also in the $\wpwmjj$ case new-physics effects due to higher-dimensional operators are most pronounced at high scales. Fig.~\ref{fig:wpwmjj-mt-eft} 
%
%
shows the impact of the dimension-eight operator $\mc{O}_{T,0}$, again
with a coefficient of $f_{T,0}/\Lambda^4=0.1$/TeV${}^4$,
supplied by the form factor of Eq.~(\ref{eq:ff}) with
$\Lambda_{F}=4.7$~TeV, on the transverse-mass distribution of the
$W^+W^-$ system compared to the SM prediction. From the high-energy
behavior of this observable we infer that the selection cut
of Eq.~(\ref{eq:wpwm-mt}) 
can efficiently improve the relative impact of the dimension-eight
operator on the SM event rate. Indeed, applying this extra cut results
in $\sigma_\mr{SM}^\mr{cut}= 3.59$~fb and $\sigma_\mr{EFT}^\mr{cut}=
4.80$~fb,  to be compared to the respective results of   
 $\sigma_\mr{SM}= 58.27$~fb and $\sigma_\mr{EFT}= 59.49$~fb 
 before the cut of Eq.~(\ref{eq:wpwm-mt}). 
While the remaining fraction of QCD background does not change much with $M_{T_{WW}}$, the $t\bar t$~(+jets) backgrounds fall precipitously with increasing $W^+W^-$~transverse mass. This is caused by the high $p_T$ of the top quarks which is needed to give the $W$~bosons sufficient transverse mass and concomitant high $p_T$. The higher top-quark transverse momentum results, however, in harder bottom quarks, which increasingly fail the central jet veto cut of $p_{T,b}<50$~GeV. One finds, therefore, that the interesting very high transverse mass region has a rather small top-pair background and, thus, is particularly well suited for restricting anomalous quartic couplings at an FCC.

%
\section{Summary and conclusions}
\label{sec:summ}
%
%
%
\begin{center}
\begin{table}
\begin{center}
\begin{tabular}{|c|c|c|c|c|c|}
\hline
VBS channel &$\sigma_S$  [fb] & $\sigma_B$ [fb]
& $S/B$ &  $S/\sqrt{B}$ &  $S/\sqrt{S+B}$\\
\hline\hline
$ \wpwpjj   $    &0.48 & 0.04 & 12&  416& 115 \\
%
\hline
$ \wpzjj   $   & 0.047 & 0.008&  5.9 & 
 91   & 35 \\
%
\hline
$\zzjj    $     &0.11& 0.008& 13.7 &  213 
& 56  \\
%
\hline
$\wpwmjj $   &3.59& 4.62& 0.78   &    289 & 217   \\
\hline
\end{tabular}
\caption{
\label{tab:significances-sm}
\it
Cross sections and different  signal-background ratios for various VBS
processes in the framework of the SM  after the optimized cuts of
Secs.~\ref{sec:wpwpjj}--\ref{sec:wpwmjj} are imposed. Decays of the
weak bosons into a specific leptonic final state are included as
detailed in the text. In each case $S$ denotes the number of events
for the EW $VVjj$ production mode, while $B$ includes the number of
background events due to the respective QCD $VVjj$ channels and (in
the $\wpwmjj$ case) top-induced production modes.  An integrated
luminosity of 30~ab$^{-1}$ is assumed. Statistical errors are at the
permille level in each case. }
\end{center}
\end{table}
\end{center}
%
%
%
%
%
\begin{center}
\begin{table}
\begin{center}
\begin{tabular}{|c|c|c|c|c|c|}
\hline
VBS channel &$\sigma_S$  [fb] 
& $\sigma_B$ [fb] & $S/B$ &  $S/\sqrt{B}$ 
&  $S/\sqrt{S+B}$\\
\hline\hline
$ \wpwpjj   $    &0.66 & 0.52 & 1.27 & 159 & 105 \\
%
\hline
$ \wpzjj   $   & 0.061 & 0.055 & 1.11 & 45    & 31 \\
%
\hline
$\zzjj    $     &0.22& 0.12& 1.83 &   110
& 65  \\
%
\hline
$\wpwmjj $   &4.8& 8.2& 0.58  &    290  & 231   \\
\hline
\end{tabular}
\caption{
\label{tab:significances-eft}
\it
Cross sections and different  signal-background ratios for various VBS
processes in the framework of the SM and including the impact of the
dimension-eight operator  $\mc{O}_{T,0}$ with  $f_{T,0}/\Lambda^4=0.1$/TeV${}^4$  and the form factor of
Eq.~(\ref{eq:ff}) after the optimized cuts of
Secs.~\ref{sec:wpwpjj}--\ref{sec:wpwmjj} are imposed. Decays of the
weak bosons into a specific leptonic final state are included as
detailed in the text. In each case $S$ denotes the number of events in
the EFT scenario, while $B$ includes the number of background events
due to the respective SM EW and QCD $VVjj$ channels 
as well as (in the $\wpwmjj$ case) top-induced production modes.  An integrated
luminosity of 30~ab$^{-1}$ is assumed. Statistical errors are at the
permille level in each case.  }
\end{center}
\end{table}
\end{center}
%
%
In this article, we provided a review of VBS processes at a future
100~TeV hadron collider. We discussed how cuts tailored to specific
final states can be used to efficiently suppress QCD and top-quark
induced background contributions that exhibit entirely different
kinematic properties than the signal processes of interest.  In order
to generically explore the capability of an FCC to identify signatures
of new physics in the weak gauge boson sector, we provided
representative results obtained in an EFT approach. While they hardly
affect SM results at low scales, the considered BSM contributions have
particular impact in the high-energy regime, significantly modifying
the tails of transverse-momentum and invariant-mass
distributions. These kinematic regimes are difficult to access at the
LHC that is currently limited to a collider energy of 13~TeV. The
possibility of a future 100~TeV hadron collider thus represents
entirely new means for exploring new physics in the weak sector.

In Tabs.~\ref{tab:significances-sm} and
\ref{tab:significances-eft} we summarise signal and background cross
sections for all discussed production modes within the optimized
selection cuts we devised. For a representative integrated luminosity
of 30~ab$^{-1}$ we also list various ratios of signal and background
event numbers. We first consider the SM EW $VVjj$ production modes as
signal and the respective QCD $VVjj$ as well as, in
the $\wpwmjj$ case, top-induced production modes as
background.  When we consider the EFT prediction as
signal the sum of EW and QCD induced production cross sections is
taken as background. In the latter case we restrict ourselves to the
scenario with a non-vanishing dimension-eight operator $\mc{O}_{T,0}$
which has the largest impact on the considered observables. 

Starting from Tab.~\ref{tab:significances-eft} one can perform a very rough extrapolation to smaller Wilson coefficients of $\mc{O}_{T,0}$: differential cross sections at high scales, for fixed form factor, scale like the square of the Wilson coefficient. The significances in the last column of Tab.~\ref{tab:significances-eft}  then suggest an FCC reach of at least 0.01~TeV$^{-4}$ for the $\mc{O}_{T,0}$ coefficient. This estimate is conservative since higher form factor scales are allowed for smaller anomalous quartic gauge couplings (aQGCs) and, also, more stringent cuts, focusing on even higher 
$m_{VV}$ regions, will improve sensitivity. Decay distributions of the $W$ and $Z$~bosons can be used to  improve sensitivity and to distinguish different dimension~8 operators. Finally, for the very hard weak bosons in the sensitive region of phase space, the analysis of hadronic $V=W,Z$ decays with highly boosted $V\to q\bar q$ jets will enhance sensitivity even further. Similar improvements are expected for other operators describing aQGCs. A realistic sensitivity estimate to aQGCs at an FCC needs to take these improvements into account and, thus, would go significantly beyond the scope of the present analysis.

%
\section*{Acknowledgements}
%

We are grateful for valuable discussions to Matthias Kesenheimer. The
work of B.~J.\  and L.~S.\  has been supported in part by the
Institutional Strategy of the University of T\"ubingen (DFG, ZUK 63),
and in part by the German Federal Ministry for Education and Research
(BMBF) under grant number 05H15VTCAA. 
The work of  M.~W.\ was supported by the DFG under Grant No.\ WO 1900/2 and by the BMBF under grant number 05H15PACC1. 
The work of D.~Z.\ was supported by the BMBF under grant number 05H15VKCCA. 
The authors acknowledge support by the state of Baden-W\"urttemberg
through bwHPC and the German Research Foundation (DFG) through grant
no INST 39/963-1 FUGG.



\begin{thebibliography}{99}

\bibitem{Aad:2012tfa}
  G.~Aad {\it et al.} [ATLAS Collaboration],
  {\it "Observation of a new particle in the search for the Standard
    Model Higgs boson with the ATLAS detector at the LHC''},
  Phys.\ Lett.\ B {\bf 716} (2012) 1 
    [arXiv:1207.7214 [hep-ex]].
  
 \bibitem{Chatrchyan:2012xdj}
  S.~Chatrchyan {\it et al.} [CMS Collaboration],
  {\it "Observation of a new boson at a mass of 125 GeV with the CMS
    experiment at the LHC''},
  Phys.\ Lett.\ B {\bf 716} (2012) 30 
  [arXiv:1207.7235 [hep-ex]].
  
\bibitem{Dolan:2012rv}
  M.~J.~Dolan, C.~Englert and M.~Spannowsky,
  {\it ``Higgs self-coupling measurements at the LHC''}, 
  JHEP {\bf 1210} (2012) 112  
  [arXiv:1206.5001 [hep-ph]].
  
\bibitem{Dolan:2015zja}
  M.~J.~Dolan, C.~Englert, N.~Greiner, K.~Nordstrom and M.~Spannowsky,
  {\it ``$hhjj$ production at the LHC''}, 
  Eur.\ Phys.\ J.\ C {\bf 75} (2015) no.8,  387
  [arXiv:1506.08008 [hep-ph]].

  
\bibitem{Baglio:2015wcg}
  J.~Baglio, A.~Djouadi and J.~Quevillon,
  {\it "Prospects for Higgs physics at energies up to 100 TeV''},
  Rept.\ Prog.\ Phys.\  {\bf 79} (2016) 116201
   [arXiv:1511.07853 [hep-ph]].
 
 \bibitem{Baglio:2012np}
  J.~Baglio, A.~Djouadi, R.~Gr{\"o}ber, M.~M.~M{\"u}hlleitner,
  J.~Quevillon and M.~Spira,
  {\it "The measurement of the Higgs self-coupling at the LHC: theoretical status''},
  JHEP {\bf 1304} (2013) 151
    [arXiv:1212.5581 [hep-ph]].
   
\bibitem{Plehn:2005nk}
  T.~Plehn and M.~Rauch,
  {\it "The quartic higgs coupling at hadron colliders''},
  Phys.\ Rev.\ D {\bf 72} (2005) 053008
    [hep-ph/0507321].

\bibitem{Djouadi:1999rca}
  A.~Djouadi, W.~Kilian, M.~M\"uhlleitner and P.~M.~Zerwas,
  {\it "Production of neutral Higgs boson pairs at LHC''},
  Eur.\ Phys.\ J.\ C {\bf 10} (1999) 45
    [hep-ph/9904287].

\bibitem{Baur:2002qd}
  U.~Baur, T.~Plehn and D.~L.~Rainwater,
  {\it "Determining the Higgs boson selfcoupling at hadron colliders''},
  Phys.\ Rev.\ D {\bf 67} (2003) 033003
   [hep-ph/0211224].
 
\bibitem{Duncan:1985vj}
  M.~J.~Duncan, G.~L.~Kane and W.~W.~Repko,
  {\it "$WW$ Physics at Future Colliders''},
  Nucl.\ Phys.\ B {\bf 272} (1986) 517.
 
\bibitem{Bagger:1995mk}
  J.~Bagger, V.~D.~Barger, K.~M.~Cheung, J.~F.~Gunion, T.~Han,
  G.~A.~Ladinsky, R.~Rosenfeld and C.-P.~Yuan,
  {\it "CERN LHC analysis of the strongly interacting $WW$ system: Gold
    plated modes''},
  Phys.\ Rev.\ D {\bf 52} (1995) 3878
    [hep-ph/9504426].
 
\bibitem{Butterworth:2002tt}
  J.~M.~Butterworth, B.~E.~Cox and J.~R.~Forshaw,
  {\it "$W W$ scattering at the CERN LHC''},
  Phys.\ Rev.\ D {\bf 65} (2002) 096014
    [hep-ph/0201098].
  
 \bibitem{Aad:2014zda}
  G.~Aad {\it et al.} [ATLAS Collaboration],
  {\it "Evidence for Electroweak Production of $W^{\pm}W^{\pm}jj$ in
    $pp$ Collisions at $\sqrt{s}=8$ TeV with the ATLAS Detector''},
  Phys.\ Rev.\ Lett.\  {\bf 113} (2014) no.14,  141803
   [arXiv:1405.6241 [hep-ex]].
     
\bibitem{Khachatryan:2014sta}
  V.~Khachatryan {\it et al.} [CMS Collaboration],
  {\it "Study of vector boson scattering and search for new physics in
    events with two same-sign leptons and two jets''},
  Phys.\ Rev.\ Lett.\  {\bf 114} (2015) no.5,  051801
    [arXiv:1410.6315 [hep-ex]].
 
\bibitem{Mangano:2016jyj}
  M.~L.~Mangano {\it et al.},
  {\it "Physics at a 100 TeV pp collider: Standard Model processes''},
  arXiv:1607.01831 [hep-ph].
  
\bibitem{Goncalves:2017gzy}
  D.~Goncalves, T.~Plehn and J.~M.~Thompson,
  {\it ``Weak Boson Fusion at 100 TeV''}, 
  arXiv:1702.05098 [hep-ph].
 
\bibitem{Arnold:2008rz} 
  K.~Arnold {\it et al.},
  {\it "\textsc{Vbfnlo:} A Parton level Monte Carlo for processes with electroweak
  bosons''},
  Comput.\ Phys.\ Commun.\  {\bf 180} (2009) 1661
  [arXiv:0811.4559 [hep-ph]].

\bibitem{Czakon:2009ss}
  M.~Czakon, C.~G.~Papadopoulos and M.~Worek,
  {\it "Polarizing the Dipoles''},
  JHEP {\bf 0908} (2009) 085
    [arXiv:0905.0883 [hep-ph]].

\bibitem{Cafarella:2007pc}
  A.~Cafarella, C.~G.~Papadopoulos and M.~Worek,
  {\it ``\textsc{Helac-Phegas:} A Generator for all parton level processes''},
  Comput.\ Phys.\ Commun.\  {\bf 180} (2009) 1941
   [arXiv:0710.2427 [hep-ph]].

\bibitem{Kanaki:2000ey}
  A.~Kanaki and C.~G.~Papadopoulos,
  {\it "\textsc{Helac:} A Package to compute electroweak helicity amplitudes''},
  Comput.\ Phys.\ Commun.\  {\bf 132} (2000) 306
  [hep-ph/0002082].

\bibitem{Papadopoulos:2005ky}
  C.~G.~Papadopoulos and M.~Worek,
  {\it "Multi-parton cross sections at hadron colliders''},
  Eur.\ Phys.\ J.\ C {\bf 50} (2007) 843
   [hep-ph/0512150].

\bibitem{Bevilacqua:2011xh}
  G.~Bevilacqua, M.~Czakon, M.~V.~Garzelli, A.~van Hameren, A.~Kardos,
  C.~G.~Papadopoulos, R.~Pittau and M.~Worek,
  {\it "\textsc{Helac-Nlo}}'',
  Comput.\ Phys.\ Commun.\  {\bf 184} (2013) 986
   [arXiv:1110.1499 [hep-ph]].

\bibitem{Papadopoulos:2000tt}
  C.~G.~Papadopoulos,
  {\it "\textsc{Phegas:} A Phase space generator for automatic
    cross-section computation''},
  Comput.\ Phys.\ Commun.\  {\bf 137} (2001) 247
   [hep-ph/0007335].

\bibitem{vanHameren:2007pt}
  A.~van Hameren,
  {\it "\textsc{Parni} for importance sampling and density estimation''},
  Acta Phys.\ Polon.\ B {\bf 40} (2009) 259
  [arXiv:0710.2448 [hep-ph]].

\bibitem{vanHameren:2010gg}
  A.~van Hameren,
  {\it "\textsc{Kaleu:} A General-Purpose Parton-Level Phase Space
    Generator''},
  [arXiv:1003.4953 [hep-ph]].

\bibitem{Harland-Lang:2014zoa}
  L.~A.~Harland-Lang, A.~D.~Martin, P.~Motylinski and R.~S.~Thorne,
 {\it  "Parton distributions in the LHC era: MMHT 2014 PDFs"},
  Eur.\ Phys.\ J.\ C {\bf 75} (2015) no.5,  204
   [arXiv:1412.3989 [hep-ph]].

\bibitem{Buckley:2014ana}
  A.~Buckley, ~Ferrando, S.~Lloyd, K.~Nordstr\"om, B.~Page,
  M.~R\"ufenacht, M.~Sch\"onherr and G.~Watt,
  {\it "LHAPDF6: parton density access in the LHC precision era''},
  Eur.\ Phys.\ J.\ C {\bf 75} (2015) 132
    [arXiv:1412.7420 [hep-ph]].

\bibitem{Cacciari:2008gp}
  M.~Cacciari, G.~P.~Salam and G.~Soyez,
  {\it "The Anti-k(t) jet clustering algorithm''},
  JHEP {\bf 0804} (2008) 063
    [arXiv:0802.1189 [hep-ph]].

\bibitem{Rauch:2016pai}
  M.~Rauch,
  {\it ``Vector-Boson Fusion and Vector-Boson Scattering''}, 
  arXiv:1610.08420 [hep-ph].

\bibitem{Hagiwara:1993ck}
  K.~Hagiwara, S.~Ishihara, R.~Szalapski and D.~Zeppenfeld,
  {\it "Low-energy effects of new interactions in the electroweak boson sector''},
  Phys.\ Rev.\ D {\bf 48} (1993) 2182.

\bibitem{Degrande:2013ng}
  C.~Degrande, N.~Greiner, W.~Kilian, O.~Mattelaer, H.~Mebane, 
  T.~Stelzer, S.~Willenbrock, C.~Zhang,
  {\it "Effective Field Theory: A Modern Approach to Anomalous Couplings''},
  Annals Phys.\  {\bf 335} (2013) 21.
  [arXiv:1205.4231 [hep-ph]].

\bibitem{Eboli:2006wa}
  O.~J.~P.~Eboli, M.~C.~Gonzalez-Garcia and J.~K.~Mizukoshi,
  {\it "$p p \to j j e^\pm \mu^\pm \nu \nu$ and $j j e^\pm \mu^\mp \nu \nu$
   at ${\cal O}( \alpha_{em}^6)$ and ${\cal O}(\alpha_{em}^4 \alpha_s^2)$ for the study
   of the quartic electroweak gauge boson vertex at CERN LHC'',}
  Phys.\ Rev.\ D {\bf 74} (2006) 073005 
  [hep-ph/0606118].


\bibitem{Jager:2011ms}
  B.~J\"ager and G.~Zanderighi,
  {\it "NLO corrections to electroweak and QCD production of $W^+W^+$ plus
  two jets in the \textsc{PowhegBox}''},
  JHEP {\bf 1111} (2011) 055
  [arXiv:1108.0864 [hep-ph]].

\bibitem{unitarity-diplom}  
B.~Feigl, diploma thesis, Karlsruhe Institute of Technology (2010); 
O.~Schlimpert, diploma thesis, Karlsruhe Institute of Technology (2013). 
 
\bibitem{unitarity-calc}  
  \texttt{https://www.itp.kit.edu/vbfnlo/wiki/doku.php?id=download:form-factor}


\bibitem{Englert:2008tn}
  C.~Englert, B.~J\"ager, M.~Worek and D.~Zeppenfeld,
  {\it "Observing Strongly Interacting Vector Boson Systems at the
    CERN Large Hadron Collider''},
  Phys.\ Rev.\ D {\bf 80} (2009) 035027
  [arXiv:0810.4861 [hep-ph]].

\bibitem{Bevilacqua:2010qb}
  G.~Bevilacqua, M.~Czakon, A.~van Hameren, C.~G.~Papadopoulos and M.~Worek,
  {\it "Complete off-shell effects in top quark pair hadroproduction with
  leptonic decay at next-to-leading order''},
  JHEP {\bf 1102} (2011) 083
  [arXiv:1012.4230 [hep-ph]].

\bibitem{Bevilacqua:2015qha}
  G.~Bevilacqua, H.~B.~Hartanto, M.~Kraus and M.~Worek,
  {\it "Top Quark Pair Production in Association with a Jet with
  Next-to-Leading-Order QCD Off-Shell Effects at the Large Hadron
  Collider''},
  Phys.\ Rev.\ Lett.\  {\bf 116} (2016) no.5,  052003
   [arXiv:1509.09242 [hep-ph]].

\bibitem{Bevilacqua:2016jfk}
  G.~Bevilacqua, H.~B.~Hartanto, M.~Kraus and M.~Worek,
  {\it "Off-shell Top Quarks with One Jet at the LHC: A comprehensive
  analysis at NLO QCD''},
  JHEP {\bf 1611} (2016) 098
  [arXiv:1609.01659 [hep-ph]].

\bibitem{Weiser:2006md}
  C.~Weiser,
  {\it "A combined secondary vertex based B-tagging algorithm in CMS''},
  \texttt{CERN-CMS-NOTE-2006-014}.

\bibitem{Rainwater:1999sd}
  D.~L.~Rainwater and D.~Zeppenfeld,
  {\it "Observing $H\to W^*W^* \to e^\pm \mu\mp \not{p}_T$ in weak
    boson fusion with dual forward jet tagging at the CERN LHC''},
  Phys.\ Rev.\ D {\bf 60} (1999) 113004
   Erratum: [Phys.\ Rev.\ D {\bf 61} (2000) 099901]
  [hep-ph/9906218].

\bibitem{Kauer:2000hi}
  N.~Kauer, T.~Plehn, D.~L.~Rainwater and D.~Zeppenfeld,
  {\it "$H\to W^+W^-$ as the discovery mode for a light Higgs boson''},
  Phys.\ Lett.\ B {\bf 503} (2001) 113
  [hep-ph/0012351].
  
\bibitem{Figy:2007kv}
  T.~Figy, V.~Hankele and D.~Zeppenfeld,
  {\it "Next-to-leading order QCD corrections to Higgs plus three jet
    production in vector-boson fusion''},
  JHEP {\bf 0802} (2008) 076
  [arXiv:0710.5621 [hep-ph]].
  
 \bibitem{Campanario:2013fsa}
  F.~Campanario, T.~M.~Figy, S.~Pl{\"a}tzer and M.~Sj{\"o}dahl,
  {\it "Electroweak Higgs Boson Plus Three Jet Production at
    Next-to-Leading-Order QCD''},
  Phys.\ Rev.\ Lett.\  {\bf 111} (2013) no.21,  211802
    [arXiv:1308.2932 [hep-ph]].
  
 \bibitem{Jager:2014vna}
  B.~J\"ager, F.~Schissler and D.~Zeppenfeld,
  {\it "Parton-shower effects on Higgs boson production via
    vector-boson fusion in association with three jets''},
  JHEP {\bf 1407} (2014) 125
    [arXiv:1405.6950 [hep-ph]].
  


\end{thebibliography}
\end{document}